\documentclass[aps,prl,reprint,superscriptaddress,nofootinbib]{revtex4-2}

\usepackage{graphicx ,subfigure}
\usepackage{amsmath,amssymb}
\usepackage{mathtools}
\usepackage{multirow}
\usepackage{textcomp}
\usepackage{float}
\usepackage{xcolor}
\usepackage{ulem}
\usepackage{dsfont}
\usepackage{bibunits}
\usepackage{import}
\usepackage{braket}
\usepackage{comment}
\usepackage{bm}
\usepackage{bbm}
\usepackage[german,english]{babel}
\usepackage[colorlinks=true,linkcolor=blue,citecolor=blue,breaklinks]{hyperref}

\renewcommand{\vec}[1]{\ensuremath{\mathbf{#1}}}

\let\oldsfdefault\sfdefault
\usepackage[scaled=0.92]{helvet}
\renewcommand{\sfdefault}{\oldsfdefault}

\defaultbibliographystyle{apsrev4-2}

\begin{document}
\normalem

\title{Thouless Pumps and Bulk-Boundary Correspondence \\ in Higher-Order Symmetry-Protected Topological Phases}

\author{Julian F. Wienand}
\address{Department of Physics and Arnold Sommerfeld Center for Theoretical Physics (ASC), Ludwig-Maximilians-Universit\"at M\"unchen, Theresienstr. 37, D-80333 Munich, Germany}
\address{Munich Center for Quantum Science and Technology (MCQST), Schellingstr. 4, D-80333 Munich, Germany}
\address{Max-Planck-Institut f\"ur Quantenoptik, 
             Hans-Kopfermann-Strasse 1, D-85748 Garching, Germany}

\author{Friederike Horn}
\address{Department of Physics and Arnold Sommerfeld Center for Theoretical Physics (ASC), Ludwig-Maximilians-Universit\"at M\"unchen, Theresienstr. 37, D-80333 Munich, Germany}
\author{Monika Aidelsburger}
\address{Department of Physics and Arnold Sommerfeld Center for Theoretical Physics (ASC), Ludwig-Maximilians-Universit\"at M\"unchen, Theresienstr. 37, D-80333 Munich, Germany}
\address{Munich Center for Quantum Science and Technology (MCQST), Schellingstr. 4, D-80333 Munich, Germany}

\author{Julian Bibo}
\address{Munich Center for Quantum Science and Technology (MCQST), Schellingstr. 4, D-80333 Munich, Germany}
\address{Department of Physics, T42, Technical University of Munich, D-85748 Garching, Germany}

\author{Fabian Grusdt}
\address{Department of Physics and Arnold Sommerfeld Center for Theoretical Physics (ASC), Ludwig-Maximilians-Universit\"at M\"unchen, Theresienstr. 37, D-80333 Munich, Germany}
\address{Munich Center for Quantum Science and Technology (MCQST), Schellingstr. 4, D-80333 Munich, Germany}

\begin{bibunit}

\begin{abstract}
The bulk-boundary correspondence relates quantized edge states to bulk topological invariants in topological phases of matter. In one-dimensional symmetry-protected topological systems (SPTs), quantized topological Thouless pumps directly reveal this principle and provide a sound mathematical foundation. Symmetry-protected higher-order topological phases of matter (HOSPTs) also feature a bulk-boundary correspondence, but its connection to quantized charge transport remains elusive. Here we show that quantized Thouless pumps connecting $C_4$-symmetric HOSPTs can be described by a tuple of four Chern numbers that measure quantized bulk charge transport in a direction-dependent fashion. Moreover, this tuple of Chern numbers allows to predict the sign and value of fractional corner charges in the HOSPTs. We show that the topologically non-trivial phase can be characterized by both quadrupole and dipole configurations, shedding new light on current debates about the multi-pole nature of the HOSPT bulk. By employing corner-periodic boundary conditions, we generalize Restas's theory to HOSPTs. Our approach provides a simple framework for understanding topological invariants of general HOSPTs and paves the way for an in-depth description of future dynamical experiments.
\end{abstract}
\date{\today}

\maketitle

\emph{Introduction.--}
Protected edge states are a signature phenomenon in (many-body) quantum systems with non-trivial topology. In one dimension (1D), such accumulation of charge at the boundary can be understood as the consequence of polarization in the bulk. As discovered by King-Smith and Vanderbilt \cite{KingSmith1993}, the polarization is a manifestation of the Zak (Berry) phase of the underlying Bloch bands \cite{Zak1989, Atala2013}. For interacting many-body systems with periodic boundaries this result was later generalized by Resta, who related polarization to the many-body position operator in 1D \cite{Resta1998}. The underlying intuition is that building up polarization in the bulk or charge at the boundary requires quantized charge transport as described by topological Thouless pumps \cite{Thouless1983,Niu1984}.

With the recent discovery of higher-order topological insulators (HOTIs) \cite{Benalcazar2017, Benalcazar2017a}, efforts were made to generalize these concepts to describe electrical multi-pole moments \cite{Benalcazar2017, Benalcazar2017a, Kang2019, Watanabe2020, Ono2019, Ren2021, Wheeler2019, Kang2021} and higher-order Thouless pumps \cite{Benalcazar2020, Benalcazar2017a, Kang2019, Petrides2020, Kang2021}. A $n$-dimensional bulk with topology of order $m$ can exhibit ($n-m$)-dimensional corner or hinge states, when open boundary conditions (OBC) are applied. Such systems have been realized in solids and classical meta-materials \cite{Benalcazar2020, Noguchi2021,Peterson2018, Imhof2018, SerraGarcia2018, Bao2019, Mittal2019, Ni2019, Ni2020, Xue2018, Dutt2020}. Higher-order boundary states are anticipated to have versatile applications in electronics and photonics \cite{Xie2021}, e.g. for topological nano-lasers \cite{Zhang2021, Kim2020}.  

Higher-order topological invariants have been proposed for both band insulators in a single-particle picture (HOTIs) and interacting quantum many-body systems (HOSPTs) protected by crystalline symmetries \cite{Benalcazar2017a, Fukui2018, Wang2018, Kang2021, Petrides2020,Araki2020a, You2020, Wheeler2019,Kang2019, You2018, Dubinkin2019, Rasmussen2020, Guo2021}. Yet, there is an ongoing debate on which of the proposed quantities constitute true bulk invariants and how exactly the multipole polarization can be calculated in extended systems with periodic boundary conditions \cite{Ono2019, Wheeler2019}. 
For instance, recent works \cite{Wheeler2019, Kang2019} proposed to extend the work of Resta, that connects the polarization to the Zak (Berry) phase \cite{Resta1998}, by defining a many-body quadrupole operator. However, these approaches have sparked controversy \cite{Ono2019}. 

In this letter, we provide a theoretical framework for understanding bulk polarization in HOSPT phases of matter. By introducing corner-periodic boundary conditions (CPBC) we extend Resta's argument \cite{Resta1998} to higher-order systems. This allows us to describe charge transport between corners during Thouless pumping cycles in a direction dependent fashion. Moreover, charge flow can be precisely tracked and an intuitive picture of bulk polarization in HOSPTs emerges. 

Our results show that quantized Thouless pumps connecting topologically distinct $C_4$-symmetric HOSPTs can be characterized by a tuple of four Chern numbers.
The underlying Zak (Berry) phases are quantized in the $C_4\times\mathbb{Z}_2$-symmetric HOSPT phases and serve as topological invariants of the latter. The invariants we define are similar to those introduced by Araki et al. \cite{Araki2020a}, but without the necessity to introduce magnetic flux in the bulk -- hence they yield a definite value for \emph{any} gapped phase in the thermodynamic limit. Our approach allows to directly relate the quantized corner charge in an HOSPT \cite{Ren2021, Watanabe2020, Bibo2020} to the Zak (Berry) phase, giving new physical meaning to the latter.

\begin{figure}[t!]
\includegraphics[width=1\columnwidth]{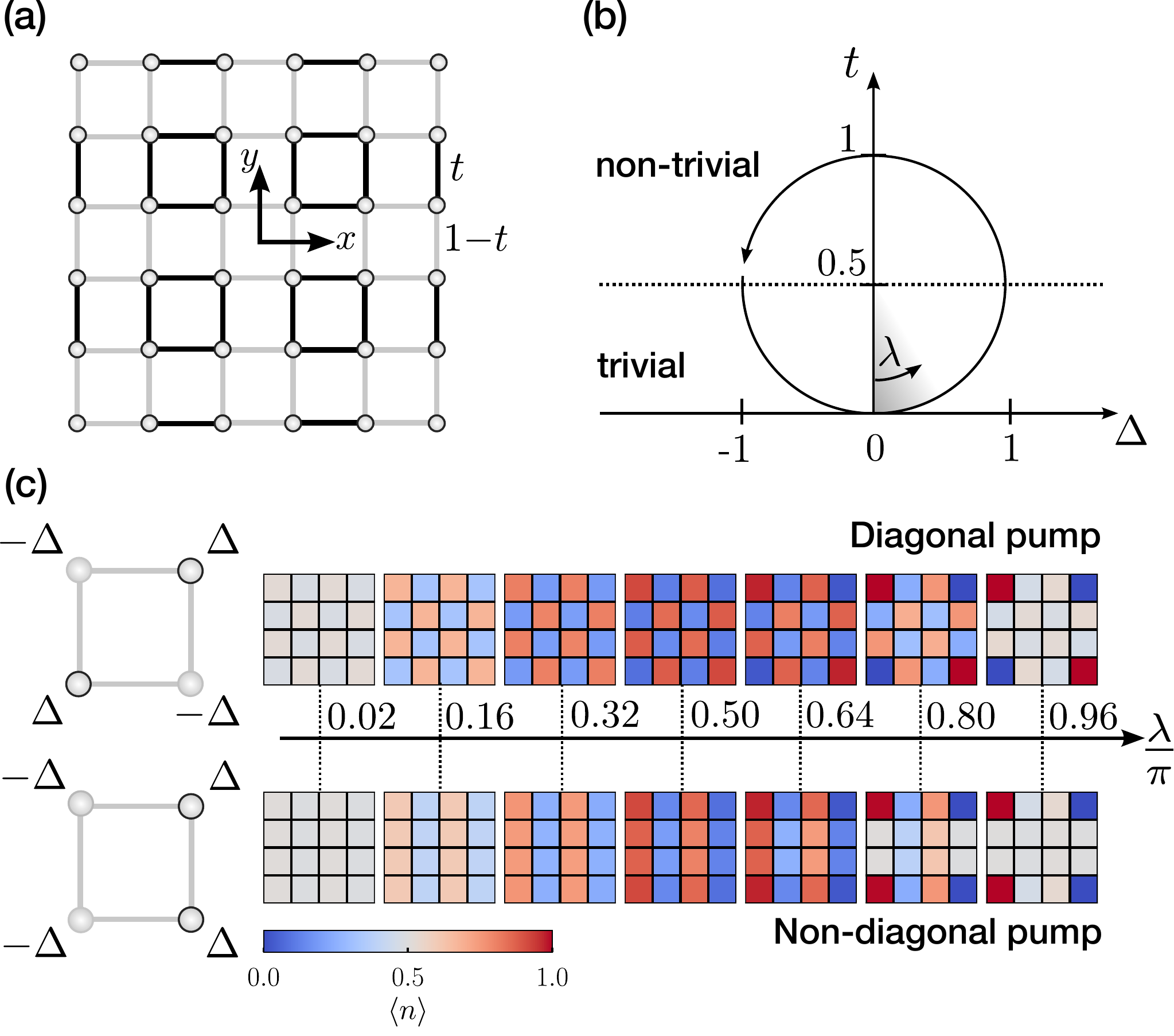}
\caption{\textbf{Thouless pumps in the 2D SL-BHM.} (a) Interacting bosons on a square lattice with staggered tunneling (strengths $t$ and $1-t$, respectively) and OBC. (b) Thouless pump, parametrized by $\lambda \in [0, 2 \pi)$, as defined in the text, with $\Delta$ controlling additional on-site potentials shown in (c). (c) Density evolution during a diagonal (upper panel) and a non-diagonal (lower panel) half Thouless pump, ending in a quadrupole and a dipole configuration, respectively.  On the left the corresponding arrangements of the additional on-site potentials are sketched: For creating a diagonal Thouless pump, in each plaquette shifts of equal sign are added on diagonally opposite sites. For creating a non-diagonal Thouless pump, shifts of the same sign are added on the same side.
} 
\label{fig:density_evol}
\end{figure}

For concreteness, we discuss interacting bosonic $C_4\left(\times\mathbb{Z}_2\right)$-symmetric HOSPTs. For these systems, we propose higher-order Thouless pumps which are in reach of current experiments with ultracold atoms and classical meta-materials \cite{Lohse2016, Lohse2018, Benalcazar2020, Lu2016, Nakajima2016}. We show that different types of pumps can create non-trivial HOSPTs in a quadrupole configuration (with vanishing dipole) and a dipole configuration (with vanishing quadrupole), see Fig.~\ref{fig:density_evol}.

\emph{Model.--}
As a hallmark model exhibiting a higher-order symmetry-protected phase, we study the 2D  superlattice-Bose-Hubbard  model (SL-BHM) \cite{Bibo2020, You2020, Grusdt2013}, which can be experimentally realized using ultracold atoms \cite{Aidelsburger2011, Nascimbene2012, Aidelsburger2013, Dai2017}. On a square lattice with OBC it is defined by the Hamiltonian
\begin{multline}
    \hat{H}^{\rm{OBC}} =- \Big[ \sum_{x=-D}^{D-1} \sum_{y=-D}^{D} \left( t(x)\, \hat{a}^{\dagger}_{x,y} \hat{a}_{x+1,y} + \mathrm{h.c.} \right) + x \leftrightarrow y \Big] \\+ \frac{U}{2} \sum_{x,y=-D}^{D} \hat{n}_{x,y} (\hat{n}_{x,y}-1),
    \label{eq:hamil}
\end{multline}
where $D=(L-1)/2$, $\hat{a}^{\dagger}_{x,y} (\hat{a}_{x,y})$ is the creation (annihilation) operator at site $(x,y)$, $\hat{n}_{x,y} = \hat{a}^{\dagger}_{x,y} \hat{a}_{x,y}$ is the particle number operator and $U$ is the on-site interaction energy, see Fig.~\ref{fig:density_evol}(a). The origin $(0, 0)$ is the $C_4$-symmetry center and for $U\to\infty$ the model has an additional $\mathbb{Z}_2$-symmetry, $\hat{a}^{\dagger}_{x,y} \leftrightarrow \hat{a}_{x,y}$. The hopping amplitudes $t(\zeta), \zeta \in \{x,y\}$ are staggered:
\begin{equation}
t(\zeta)  = \begin{cases}
1-t &\mathrm{for} \; \zeta \in  \{-D,-D+2,\dots,D-1\}\\
t & \mathrm{for}  \;\zeta \in  \{-D+1,-D+3,\dots,D-2\}
\end{cases}
\end{equation}
with $t \in [0,1]$ controlling the transition from the trivial ($t=0$) to the topological ($t=1$) phase \cite{Bibo2020}. In the following, we propose two types of Thouless pumps in this model.

\emph{Thouless pumping cycle.--}
A Thouless pump is the cyclic adiabatic variation of an external parameter. It leads to quantized charge transport that characterizes the topology of the bulk \cite{Thouless1983, Thouless1982Chern}. For the 2D SL-BHM our full pumping cycle consists of a closed trajectory in a $\Delta$-$t$ parameter space. It crosses two $C_4$-symmetric points and avoids closing the bulk gap, see Fig.~\ref{fig:density_evol}(b). Here, $\Delta$ controls the strength of additional on-site potentials whose arrangement dictates the direction of the charge transport. We will show two types of Thouless pumps that transport charge diagonally (diagonal pump) or horizontally (non-diagonal pump). For the former, each plaquette has on-site potentials in a cross-diagonal arrangement, Fig.~\ref{fig:density_evol}(c) top left; for the latter, each plaquette has on-site potentials with equal sign on the same side, Fig.~\ref{fig:density_evol}(c) bottom left.
The total Hamiltonians then read:
\begin{equation}
\begin{aligned}
    &\hat{H}^{\rm{diag.}} = \hat{H}^{\rm{OBC}} + \Delta \sum_{x,y=-D}^{D} \hat{n}_{x,y} (-1)^{(x+D)+(y+D)}\\
    &\hat{H}^{\rm{non-diag.}} = \hat{H}^{\rm{OBC}} - \Delta \sum_{x,y=-D}^{D} \hat{n}_{x,y} (-1)^{(x+D)}.
    \label{eqHpumpDef}
\end{aligned}
\end{equation}
The pump cycle is parametrized by $\lambda\in[0,2\pi)$, with $t(\lambda)=(1+\cos(\lambda)) / 2$ and $\Delta(\lambda)=\sin(\lambda)$,  as illustrated in Fig.~\ref{fig:density_evol}(b). It breaks the $C_4$-symmetry, except for $\lambda \in \pi \mathbb{Z}$. 

In Fig.~\ref{fig:density_evol}(c) we show the density evolution of the diagonal (upper panel) and non-diagonal (lower panel) pumps with OBC and at half-filling ($N=L^2/2$). We use exact diagonalization for $L=4$ and assume hard-core bosons, i.e. $U \to \infty$.
At the beginning of the pump, charge accumulates at the sites which are subject to negative energy shifts $-\Delta$. Then, once $\lambda=\pi/2$ is passed and $|\Delta|$ decreases, the density evens out in the bulk and along the edges. At the corners, however, the average density increases further up to $1$ or down to $0$, respectively, until $\lambda = \pi$. This yields four corner-localized fractional charges, two with charge $-1/2$ and two with charge $+1/2$. The arrangement of these corner charges at $\lambda = \pi$ corresponds either to a quadrupole (diagonal pump) or a dipole (non-diagonal pump) configuration.

\emph{Higher-order Zak phase and bulk-boundary correspondence in HOSPTs.--}
Next, we develop a theoretical framework relating the fractional corner charges of the HOSPTs at $\lambda = \pi \mathbb{Z}$ to bulk properties. By introducing CPBC we define a tuple of Zak (Berry) phases that act as topological invariants for HOSPTs. In addition, each Zak (Berry) phase will be associated with a certain direction, such that its change can be connected to a current operator pointing along that direction.

\begin{figure}[t!]
\includegraphics[width=1\columnwidth]{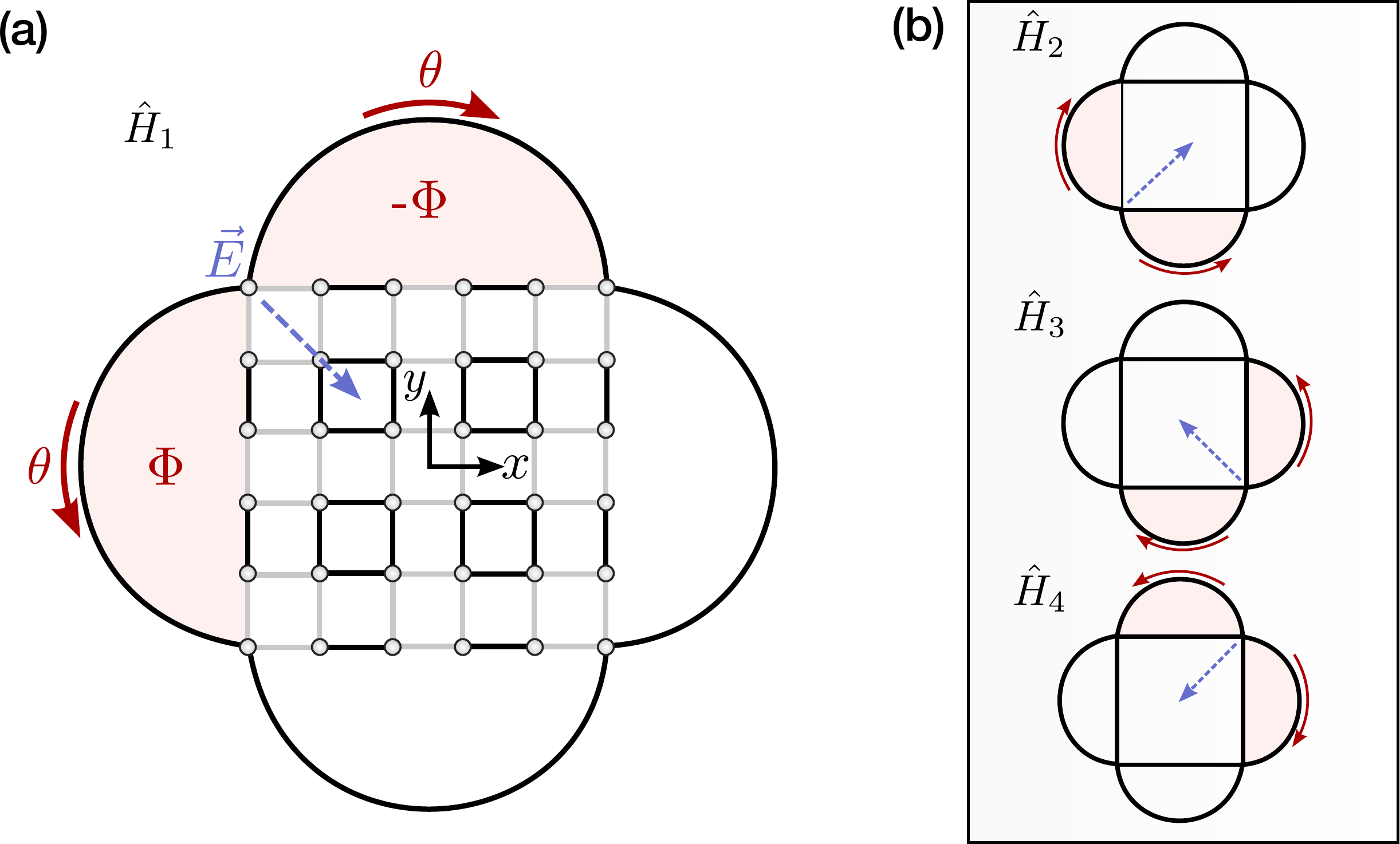}
\caption{\textbf{Direction-dependent sensing of charge flow.} Model with corner-periodic boundary conditions (CPBC). We propose four gauge choices $\hat{U}_1$ (a), $\hat{U}_2$, $\hat{U}_3$, $\hat{U}_4$ (b) that are connected by $C_4$-symmetry, Eq.~\eqref{eq:gauge_hamil}. The Zak (Berry) phases defined on these choices act as sensors of charge flow. They are only sensitive in the direction of the electric field (blue) that is induced when the flux in the outer super-cells becomes time-dependent.} 
\label{fig:gauge_choice}
\end{figure}

To achieve CPBC, as illustrated in Fig.~\ref{fig:gauge_choice}(a), we add corner-connecting links to the Hamiltonian in Eq.~\eqref{eq:hamil}:
\begin{equation}
    \hat{H}^{\rm{C}} = -t\left( \hat{a}^{\dagger}_{c_{1}} \hat{a}_{c_{2}} +
    \hat{a}^{\dagger}_{c_{2}} \hat{a}_{c_{3}} 
    + \hat{a}^{\dagger}_{c_{3}} \hat{a}_{c_{4}} +
    \hat{a}^{\dagger}_{c_{4}} \hat{a}_{c_{1}}\right),
\end{equation}
where $c_{i}$ denotes the coordinates of the $i$-th corner, i.e. $c_{1}=(-D, D)$, $c_{2}=(-D, -D)$, $c_{3}=(D, -D)$ and $c_{4}=(D, D)$ with $D = L/2 -1/2$. The total Hamiltonian with CPBC then reads:
$\hat{H}^{\rm{CPBC}} = \hat{H}^{\rm{OBC}} + \hat{H}^{\rm{C}}$.
With CPBC applied, the four corner sites form one additional plaquette. They also give rise to four super-cells outside the bulk, delimited by the edge of $\hat{H}^{\rm{OBC}}$ and one of the corner-connecting links, see Fig. \ref{fig:gauge_choice}(a). 

We start by extending the definition of the (many-body) Zak (Berry) phase to higher-order systems. To this end, magnetic flux $\Phi$ is adiabatically inserted in the two super-cells meeting at corner $i$. This process is associated with an induced electric field pointing along a diagonal \cite{supp}, see Fig.~\ref{fig:gauge_choice}, and can be formally described by gauge transformations $\hat{U}_i$, $i\in \{1,2,3,4\}$, which we apply \emph{only} to the corner-parts of the Hamiltonian:
\begin{equation}
  \hat{H}_{i}^{\rm{C}}(\theta) = \hat{U}^{\dagger}_i(\theta) \hat{H}^{\rm{C}} \hat{U}_{i}(\theta) ,
  \label{eq:gauge_hamil}
\end{equation}
with $\hat{U}_{i}(\theta) = e^{i\,\hat{X}_{i}(\theta)}$ and $\hat{X}_{i}(\theta) = \theta\; \hat{n}_{c_{i}}$.
Here, $\hat{n}_{c_{i}}$ is the particle number operator at the $i$-th corner. 

The four gauge transformations $\hat{U}_i$ with $i\in\{1,2,3,4\}$ are related to each other through $C_{4}$-symmetry, i.e. $C^{-1}_{4} \hat{U}_{i}(\theta) C_4 = \hat{U}_{i+1}(\theta)$. The resulting Hamiltonians, $\hat{H}_i(\theta) = \hat{H}_{i}^{\rm{C}}(\theta) + \hat{H}^{\rm{OBC}}$, are sketched in Fig.~$\ref{fig:gauge_choice}$: Each gauge transformation adds a phase $\theta$ to a pair of corner-connecting links. As desired, the super-cells outside the bulk are pierced by a flux $\Phi=\theta$. 

\begin{figure}[b!]
\includegraphics[width=1\columnwidth]{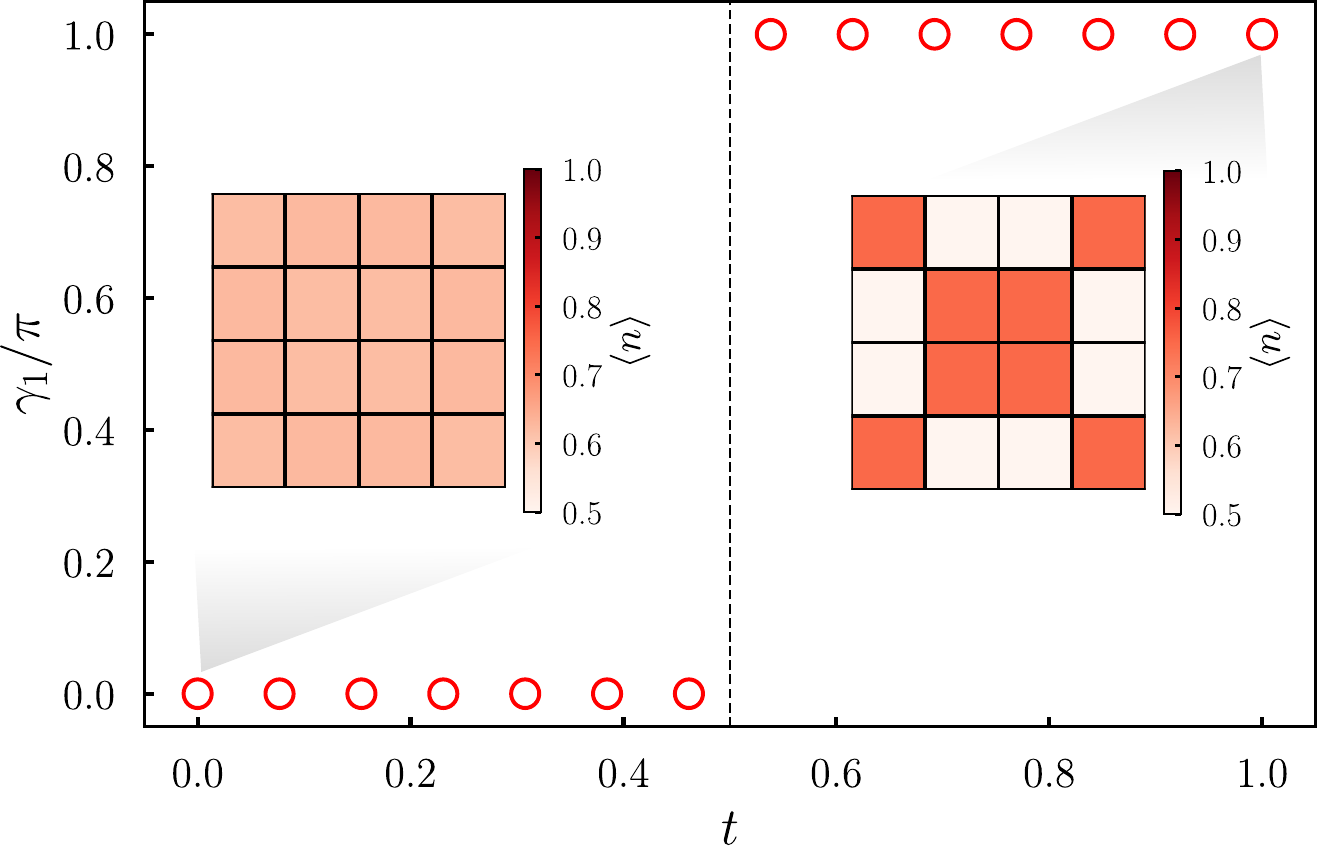}
\caption{\textbf{Quantized higher-order Zak (Berry) phase} We show $\gamma_1$ as a function of the tunneling parameter $t$ at half-filling ($N=L^2/2$) and with CPBC for $L=4$. The insets show the density expectation values of a $4\times{4}$ system at filling $N=L^2/2 +2$. Even though the non-trivial phase ($t>0.5$) does not exhibit any corner states due to CPBC, the two extra particles above half-filling lead to an occupation imbalance between edge doublets and bulk plaquettes that is unique to the non-trivial phase.
}
\label{fig:araki}
\end{figure}

Next, separately for each gauge choice $\hat{U}_i$, we define a higher-order Zak (Berry) phase~$\gamma_i$ as the geometric phase picked up by the ground state wavefunction $\ket{\psi_i(\theta)}$ of $\hat{H}_i(\theta)$ when changing $\theta$ from $0$ to $2 \pi$ \cite{Berry1984,Zak1989}:
\begin{equation}
    \gamma_i = \oint_{0}^{2 \pi} d\mathbf{\theta} ~ \bra{\psi_i(\theta)} i \partial_\theta \ket{\psi_i(\theta)}.
    \label{eq:berryphase}
\end{equation}
We note that, like the 1D Zak phase \cite{Zak1989}, the higher-order version $\gamma_i$ explicitly depends on our choice of gauge for inserting $2 \pi$ flux through the super-cells, while the difference $\Delta\gamma_i$ is gauge-invariant.

As a direct consequence of $\mathbb{Z}_2$-symmetry in the case of hard-core bosons, the higher-order Zak (Berry) phase is $\mathbb{Z}_2$-quantized,
\begin{equation}
    \gamma_i \in \pi \;\mathbb{Z},
    \label{eq:berryquant}
\end{equation}
see supplements \cite{supp} for an explicit proof.

The higher-order Zak (Berry) phases we introduce are related to the $C_4$-symmetry-protected geometric phases proposed by Araki et al.~\cite{Araki2020a, supp}.  However, in contrast to the construction in \cite{Araki2020a}, our bulk Hamiltonian $\hat{H}^{\rm{OBC}}$ remains independent of $\theta$ and our gauge choice creates a twist of the Hamiltonian $\hat{H}_{i}(\theta)$ \emph{without} introducing flux in the bulk. Hence, the gap remains open during flux insertion in the thermodynamic limit \cite{supp}, rendering Eq.~\eqref{eq:berryphase} a well-defined topological invariant. Extending Araki's scheme by introducing their fluxes through our corner-periodic links yields a robust $\mathbb{Z}_4$-quantized invariant protected by $C_4$-symmetry.

Fig.~\ref{fig:araki} depicts $\gamma_1$ as a function of the tunneling parameter $t$ in the SL-BHM at $C_4\times\mathbb{Z}_2$-symmetric points (the plots for $\gamma_{2,3,4}$ look identical). The wavefunctions in Eq.~\eqref{eq:berryphase} were calculated in a small system ($L=4$) with CPBC. The Zak (Berry) phase is quantized, as predicted, and jumps from $0$ (trivial phase) to $\pi$ (non-trivial phase). With CPBC applied, the non-trivial phase is characterized by a density imbalance between bulk and edge doublets.

Finally, we relate the higher-order Zak (Berry) phase to charge transport and derive a bulk-boundary correspondence for HOSPTs. This is achieved by extending Resta's argument to higher-order systems and introducing a many-body position operator in the bulk (see \cite{supp} Sec.~\ref{sec:resta} for details). A key step in this process is to note that the adiabatic flux insertion in Eq.~\eqref{eq:gauge_hamil} can be directly related to the current passing diagonally through a corner, $\hat{J}_{i} = \partial_\theta \hat{H}_i (\theta)|_{\theta=0}$ for $i=1,...,4$. Integrating up these currents along an adiabatic path connecting two HOSPTs (e.g., along a half-Thouless pump cycle) yields a total change of the corner charge $\Delta q_{c_i}$ in corner $i$, and we can show (\cite{supp} Sec.~\ref{sec:currentOperator}) that 
\begin{equation}
    \Delta q_{c_i} = -\frac{\Delta \gamma_i}{2 \pi}.
    \label{eqDqciResult}
\end{equation}
I.e. the fully-gauge invariant difference $\Delta \gamma_i$ of the higher-order Zak (Berry) phases in two HOSPTs is directly related to the difference of their corner charges. Since we showed that $\gamma_i$ is quantized by $C_4 \times \mathbb{Z}_2$-symmetry, it follows that the corner charge $\Delta q_{c_i}$ is also quantized and represents an intrinsic topological invariant distinguishing HOSPTs.

\emph{Chern numbers of higher-order Thouless pumps.--}
We can now apply the higher-order Zak (Berry) phase defined in Eq.~\eqref{eq:berryphase} to track the charge flow during the Thouless pumps introduced in Fig.~\ref{fig:density_evol}.
The total amount of charge $\Delta Q_{c_i} = \oint dq_{c_i}$ transported during one full pumping cycle, or equivalently, the amount of charge piling up at the corners as corner states with OBC, can be measured by four Chern numbers $\mathcal{C}_i$ with $i\in\{1,2,3,4\}$. Using our main result from Eq.~\eqref{eqDqciResult}, the latter are obtained as winding numbers of the higher-order Zak (Berry) phase \cite{signnote},
\begin{equation}
    \mathcal{C}_i = \oint_{0}^{2 \pi} \frac{\mathrm{d}\lambda}{2\pi}\, \partial_{\lambda} \gamma_{i} (\lambda) = \sum_n \left[\frac{\gamma_i(\lambda_{n+1})}{2\pi} - \frac{\gamma_i(\lambda_{n})}{2\pi}\right].
    \label{eqDefCi}
\end{equation}
The second expression is a discretized version, with a sufficiently large number of discrete points $\lambda_n \in [0,2\pi)$. Our conventions are such that a negative (positive) Chern number $\mathcal{C}_i$ indicates a particle current from the center toward the corner $c_i$ (from the the corner $c_i$ toward the center), see \cite{supp} Sec.~\ref{sec:currentOperator}.

Since the Zak (Berry) phase is defined mod $2 \pi$, it follows directly from Eq.~\eqref{eqDefCi} that the Chern numbers, $\mathcal{C}_i$, and the associated bulk charge transport along the corresponding diagonal $\Delta Q_{c_i} = -\mathcal{C}_i$, are integer quantized. Note, this remains true even at finite $U$ where $\mathbb{Z}_2$ symmetry is broken. Moreover, by $C_4$-symmetry, half-Thouless pumps connecting HOSPTs lead to a change of the corner charge $\Delta q_{c_i} = -\mathcal{C}_i/2$ given by half the Chern number. The sum-rule $\sum_{i=1}^4 \mathcal{C}_i = 0$ guarantees net charge conservation.

\begin{figure}[tbh!]\label{fig:ChernNumbers}
\begin{center}
\includegraphics[width=0.45\textwidth]{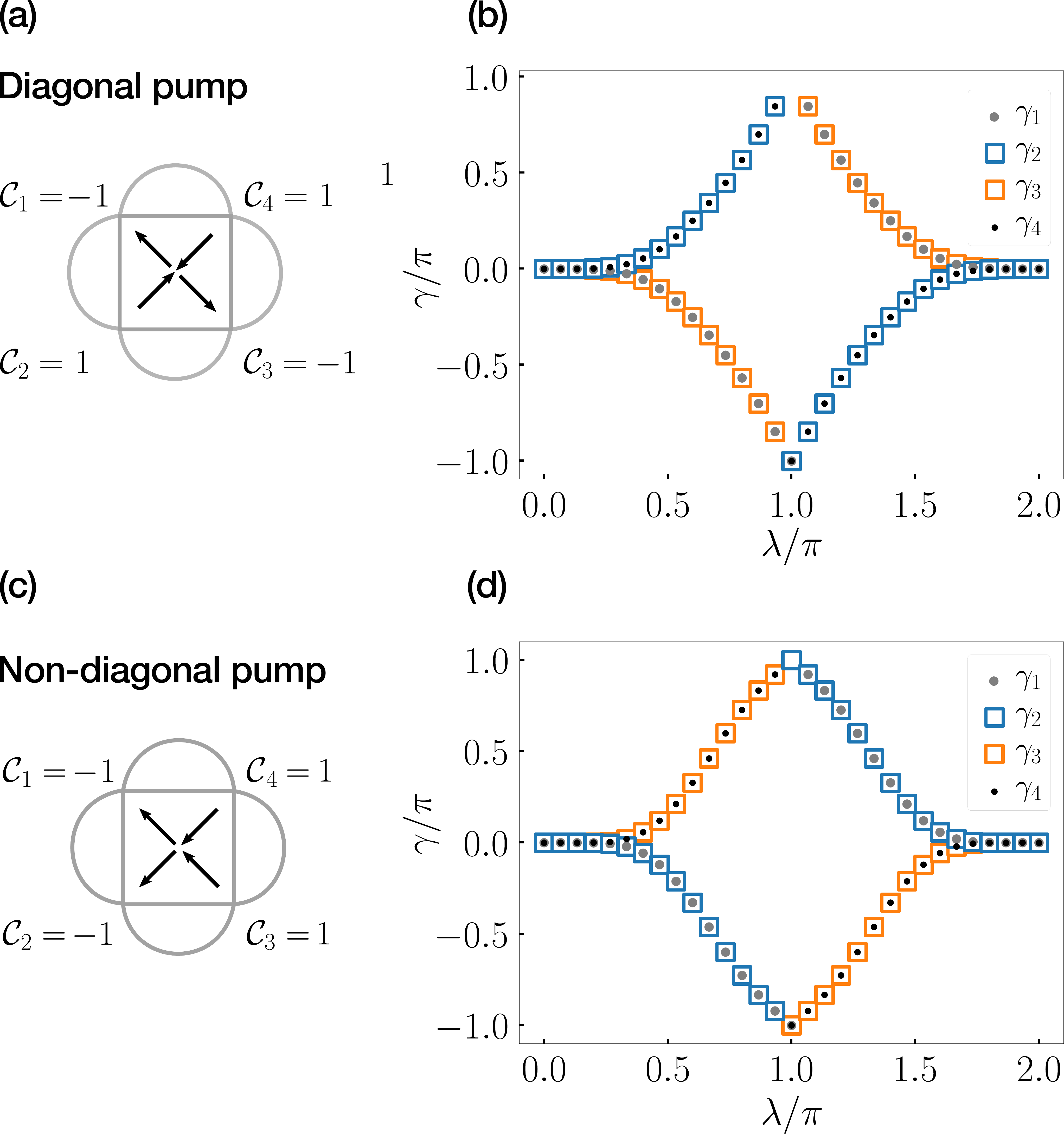}
\end{center}
\caption{\textbf{Chern number tuples for diagonal and non-diagonal Thouless pumps.} (b, d) Evolution of the higher-order Zak (Berry) phases $\gamma_i$ along the full non-diagonal (lower panel) and diagonal (upper panel) Thouless pumps in the SL-BHM, as defined below Eq.~\eqref{eqHpumpDef}. (a, c) Each Chern number $\mathcal{C}_i$ is associated with the current operator $\hat{J}_i$ defined through the gauge transformation $\hat{U}_i$. A negative (positive) value indicates charge flow to (away from) the corner $c_i$.}
\label{fig:berry_evolution}
\end{figure}

Now we calculate the Chern numbers characterizing the higher-order Thouless pumps introduced earlier for the SL-BHM. Figs.~\ref{fig:berry_evolution}(b), (d) show the evolution of the four higher-order Zak (Berry) phases as a function of the pump parameter $\lambda$. The Chern numbers $\mathcal{C}_i$ are extracted from the windings of $\gamma_i$ and read $\mathcal{C}^{\text{diag.}} = (-1, 1, -1, 1)$ for the diagonal and $\mathcal{C}^{\text{non-diag.}}=(-1, -1, +1, +1)$ for the non-diagonal pump. The resulting overall charge flow is sketched in  Figs.~\ref{fig:berry_evolution}~(a), (c) for both pumps. The result is consistent with the density evolution we find in Fig.~\ref{fig:density_evol} for a system with OBC. There, the half-charged particle (hole) corner states emerge where the associated Chern number is negative (positive) -- in accordance with  our result $\Delta q_{c_i} = -\mathcal{C}_i/2$.

Our example shows that the tuple of Chern numbers $\mathcal{C}_i$ can describe Thouless pumps building up both a dipole and a quadrupole moment. This sheds light on previously reported difficulties with defining a pure quadrupole operator in systems without dipole conservation~\cite{Ono2019}. Our case study also demonstrates that three Chern numbers need to be known to distinguish diagonal from non-diagonal pumps.

\emph{Summary and Outlook.--}
In conclusion, we have investigated quantized charge transport in higher-order topological systems and provided a description of higher-order Thouless pumps. In doing so, we introduced the higher-order Zak (Berry) phase as a new topological invariant of HOSPTs which enters in the bulk-boundary correspondence. We have found a way to extend Resta's earlier work \cite{Resta1998} to HOSPTs and relate the many-body Zak (Berry) phase to charge transport in the bulk. For a concrete system with $C_4\left(\times\mathbb{Z}_2\right)$-invariance we demonstrated that a tuple of four Chern numbers characterizes its HOSPTs and can be used to track the emergence of dipoles and quadrupoles in the system's bulk during an experimentally accessible higher-order Thouless pump.

Our approach can straightforwardly be applied to other discrete symmetries, geometries, fillings or settings without translational symmetry in the bulk. We leave detailed analysis of such cases to future work. Particularly interesting directions include the SL-BHM at quarter filling or quasicrystals constituting HOSPTs.

\begin{acknowledgments}
\section*{Acknowledgments}
We thank Frank Pollmann, Izabella Lovas, Christian Schweizer and Cesar Cabrera for insightful discussions. J.F.W. acknowledges support from the German Academic Scholarship Foundation and the Marianne-Plehn-Program. J.B. acknowledges support from the European Research Council (ERC) under the European Union’s Horizon 2020 research and innovation programme (grant agreement No. 771537). We acknowledge funding by the Deutsche Forschungsgemeinschaft (DFG, German Research Foundation) via Research Unit FOR 2414 under project number 277974659, and under Germany's Excellence Strategy -- EXC-2111 -- 390814868.
\end{acknowledgments}

\putbib[references]
\end{bibunit}


\normalem

\renewcommand{\theequation}{S\arabic{equation}}
\renewcommand{\thefigure}{S\arabic{figure}}
\newcommand{\h}{\hat{H}}
\newcommand{\U}{\hat{U}}
\newcommand{\V}{\hat{V}}
\newcommand{\Q}{\hat{X}}
\newcommand{\n}{\hat{n}}
\newcommand{\vj}{\vec{j}}
\newcommand{\va}{\bm{\alpha}}
\newcommand{\tl}{\tilde{L}}
\newcommand{\tth}{\tilde{\theta}}
\newcommand{\dv}{\Delta{\theta}}
\setcounter{equation}{0}
\setcounter{figure}{0}
\setcounter{section}{0}
\setcounter{secnumdepth}{4}

\onecolumngrid
\begin{center}
    \large \textbf{Supplemental Material}
\end{center}

\vspace{2em}

\begin{bibunit}

In this supplement we prove that the higher-order Zak (Berry) phases defined in the main text (Eq.~\ref{eq:berryphase}) are quantized by $\mathbb{Z}_2$-symmetry and hint at a generalization to $C_4$-symmetry following Araki et.~al. as well as softcore bosons, i.e., $U<\infty$. We generalize Resta's argument~\cite{Resta1998} to higher-order symmetry-protected topological phases and show how the Zak (Berry) phases are related to quantized charge transport. Moreover, we provide additional numerical evidence that the gap does not close during adiabatic flux insertion. Whenever we take the limit of \textit{hardcore bosons at half-filling}, i.e., assuming the additional $\mathbb{Z}_2$ symmetry, we state that explicitly. Otherwise, the results generally apply to arbitrary fillings and interactions $U$. We denote creation and annihilation operators for hardcore bosons at half-filling by $\hat{b}_{x,y}$ and $\hat{b}^{\dagger}_{x,y}$, respectively.

\section{Quantization of the higher-order Zak (Berry) phase}
\label{sec:quantization}

\subsection{Quantization due to $\mathbb{Z}_2$-Symmetry}

Here we show that the higher-order Zak (Berry) phase $\gamma_i=\pi\,\mathbb{Z}$ (Eq.~\eqref{eq:berryquant}) is quantized by the particle-hole $\mathbb{Z}_2$-symmetry. In the main text we used four symmetry related gauge transformations:  
\begin{equation}
    \hat{U}_i(\theta) = e^{i\hat{X}_i(\theta)},\;\hat{X}_i(\theta)=\theta \hat{n}_{c_i}
\end{equation} with $C_4: \hat{U}_i \to \hat{U}_{i+1}$ to introduce a total flux of $\Phi=\theta$ in two adjacent super-cells with CPBC \footnote{Those gauge transformations have been related by $C_4$ symmetry, although for the quantization of this higher-order Zak (Berry) phase $C_4$ is not essential.}. The transformed Hamiltonian reads
\begin{equation}
    \label{sup_eq:hthetai}
    \h^{CPBC}_i(\theta)=\h^{OBC}+\underbrace{\hat{U}^{\dagger}_i(\theta)\h^{C}\hat{U}_i(\theta)}_{=\h^{C}_i(\theta)}
\end{equation} with $\ket{\Psi_i(\theta)}$ denoting the gapped ground state. To obtain the higher-order Zak (Berry) phase $\gamma_i$ (belonging to the gauge choice $\hat{U}_i(\theta)$), we evaluated the integral:
\begin{equation}
    \gamma_i =i\int^{2\pi}_{0}\mathrm{d}\theta \braket{\Psi_{i}(\theta)|\partial_\theta|\Psi_{i}(\theta)}.
    \label{supp_eq:berryeqsupp}
\end{equation} 
Let $\hat{S}$ be the $\mathbb{Z}_2$-valued operator,
\begin{equation}
\hat{S}:\;\prod_{x,y} ( \hat{b}_{xy} +  \hat{b}^{\dagger}_{xy} ),\quad \hat{S}^2 = \mathbbm{1},\quad \hat{S}: \hat{b}_{x,y}\leftrightarrow\hat{b}^{\dagger}_{x,y},
\end{equation} then under this symmetry the twisted Hamiltonian (of hardcore bosons at half-filling) and its ground state transform as follows: $\hat{S}\hat{H}^{CPBC}_{i}(\theta)\hat{S} = \hat{H}^{CPBC}_{i}(-\theta)$ and $\hat{S}\ket{ \Psi_{i}(\theta)} = e^{i f(\theta)}\ket{ \Psi_{i}(-\theta)}$. Here, $f(\theta)$ is a $\theta$-dependent global phase. The quantization of the higher-order Zak (Berry) phase with respect to the $\mathbb{Z}_2$-symmetry can be seen from transforming Eq.~\ref{supp_eq:berryeqsupp} as follows:
\begin{align}
\label{sup_eq:z2proof}
\nonumber
     \gamma_i =& \;i\int^{2\pi}_{0}\mathrm{d}\theta \braket{\Psi_{i}(\theta) \hat{S} |\partial_\theta| \hat{S} \Psi_{i}(\theta)} \\
     \nonumber
     =& \;i\int^{2\pi}_{0}\mathrm{d}\theta \braket{\Psi_{i}(-\theta) |\partial_\theta|  \Psi_{i}(-\theta)} - \int_0^{2 \pi} d\theta \, \partial_\theta \, f(\theta)\\
     \nonumber
     =& \;i\int^{2\pi}_{0}\mathrm{d}\theta \braket{\Psi_{i}(-\theta) |\partial_\theta|  \Psi_{i}(-\theta)} + 2 \pi\,\mathbb{Z}\\
     \nonumber
     \overset{\tilde{\theta}\,=\,-\theta}{=}& \;-i\int^{0}_{- 2 \pi}\mathrm{d}\tilde{\theta} \braket{\Psi_{i}(\tilde{\theta}) |\partial_{\tilde{\theta}}|  \Psi_{i}(\tilde{\theta})} + 2 \pi\,\mathbb{Z}\\
     \nonumber
     \overset{\Psi_i(\theta + 2 \pi) = \Psi_i(\theta)}{=}& \;-i\int^{2 \pi}_{0}\mathrm{d}\tilde{\theta} \braket{\Psi_{i}(\tilde{\theta}) |\partial_{\tilde{\theta}}|  \Psi_{i}(\tilde{\theta})} + 2 \pi\,\mathbb{Z}\\
     =& -\gamma_i  + 2 \pi\,\mathbb{Z}.
\end{align}

Consequently, $2 \gamma_i = 0\;\mathrm{mod}\;2\pi$ and $\gamma_i \in \{0, \pi\}$. Hence, all four Zak (Berry) phases are $\mathbb{Z}_2$-quantized.

One can further show that, as a direct consequence of $C_4$-symmetry, all four Zak (Berry) phases are equal modulo $2\pi$:
\begin{align}
\label{sup_eq:allberriesareequal}
    \gamma_i &=i\int^{2\pi}_{0}\mathrm{d}\theta \braket{\Psi_{i}(\theta)|\partial_\theta|\Psi_{i}(\theta)}\nonumber\\
    &=i\int^{2\pi}_{0}\mathrm{d}\theta\braket{\Psi_{i}(\theta)|C^{-1}_{4}\partial_\theta C_4|\Psi_{i}(\theta)}\nonumber\\
     &=i\int^{2\pi}_{0}\mathrm{d}\theta\braket{C_4\Psi_{i}(\theta)|\partial_\theta |C_4\Psi_{i}(\theta)}\nonumber\\
     &=i\int^{2\pi}_{0}\mathrm{d}\theta \braket{\Psi_{i+1}(\theta)|\partial_\theta|\Psi_{i+1}(\theta)}-\int^{2\pi}_{0}\mathrm{d}\theta\; \partial_\theta\varphi(\theta)\nonumber\\
     &=\gamma_{i+1}\;(\text{mod}\;2\pi)
\end{align}which implies $\gamma_1=\gamma_2=\gamma_3=\gamma_4$.

\subsection{Quantization due to $C_4$-Symmetry}

In their recent work \cite{Araki2020a}, Araki et al. propose a higher-order Zak (Berry) phase that is quantized by $C_4$-symmetry. As compared to our approach, the gauge choice in \cite{Araki2020a} introduces flux in the bulk, which risks closing the gap. This can be fixed by applying their method to the corner-connecting links when CPBC are applied. We leave a detailed analysis of this to future work. One of the main reasons behind choosing our Berry phase over Araki's is that the phase emerging in Resta's perturbation theory is indeed identical to the higher-order Zak (Berry) phase, as demonstrated in the following sections.

\subsection{Softcore Bosons}
If we consider the SL-BHM with finite interactions, the additional $\mathbb{Z}_2$ symmetry is broken and the remaining symmetries are $U(1)\times C_4$ --- still distinguishing HOSPTs in the 2D SL-BHM. This implies that our $\mathbb{Z}_2$ higher-order Zak (Berry) phase is no longer quantized in individual HOSPT phases. However, the associated Chern numbers, defined as the winding number of these phases, remain unchanged. Consequently, our higher-order Zak (Berry) phase can still be used to characterize Thouless pumps of softcore bosons \textit{even though} its quantizing symmetry is broken. 
Note that, if the higher-order Zak (Berry) phase lacks quantization, the general relation between $\Delta q_{c_i}$ and $\Delta\gamma_{i}$ shown in Eq.~\eqref{eqDqciResult} does not hold anymore. However, we believe that this relation can be shown to remain valid if the aforementioned higher-order Zak (Berry) phase quantized by $C_4$ symmetry is considered. This is left for future work.

\section{Generalization of Resta's argument}
\label{sec:two}
The goal of this section is to generalize Resta's argument~\cite{Resta1998} to higher-order topological systems and \textit{higher-order symmetry-protected topological phase} (HOSPT) with \textit{corner periodic boundary conditions} (CPBC). In a seminal work Resta defined a meaningful many-body position operator for one-dimensional systems with periodic boundary conditions. From this, an expression for the electric polarization (dipole per unit length) in the thermodynamic limit $L\to\infty$ was obtained. The many-body position operator\footnote{Note, in Ref.~\cite{Resta1998} the position operator is denoted by $\hat{X}$. However, to not confuse with the generator of the gauge transformation used in this work we denoted it differently.} denoted by $\hat{R}$ and the electric polarization were defined as follows: 

\label{sec:resta}
\begin{equation}
    \braket{\hat{R}} = \frac{L}{2\pi}\text{Im}\log\braket{\Psi_0|e^{\frac{2\pi i}{L} \hat{R}}|\Psi_0},\qquad P_{el} =\lim_{L\to\infty}\frac{e}{2\pi}\text{Im}\log\braket{\Psi_0|e^{\frac{2\pi i}{L} \hat{R}}|\Psi_0}
\end{equation}where $\ket{\Psi_0}$ is the many-body ground state and $e$ the electric charge. Note that $\braket{\hat{R}}$ is defined modulo $L$ and $P_{el}$ modulo $e$. To obtain this result a one-dimensional family of Hamiltonians $\h(\Phi/L)$ with total magnetic flux $\Phi$ and PBC were considered. By inserting a $2\pi$ flux and using perturbation theory, a relation between the many-body position operator and the Zak (Berry) phase $\gamma_L$ was found:
\begin{equation}
    \braket{\hat{R}}=\frac{L}{2\pi}\gamma_L\quad\Rightarrow\quad P_{el} = \lim_{L\to\infty}\frac{e}{2\pi}\gamma_L.
\end{equation}By evaluating the time-derivative (assuming an adiabatic change of parameters) of $\braket{\hat{R}}$ it was confirmed, based on Ref.~\cite{Thouless1983}, that $P_{el}$ indeed corresponds to the electric polarization. From this, one could infer (see also Ref.~\cite{Ortiz1994}) that the change of the electric polarization over one period of time\footnote{Note that we assume $\h(\tau+T)=\h(\tau).$} $T$
\begin{equation}
  \Delta P_{el} =   \int^{T}_0\mathrm{d}\tau\; \frac{\partial}{\partial \tau}P_{el}(\tau) =  \int^{T}_0\mathrm{d}\tau\;{J(\tau)} = \lim_{L\to\infty}\frac{e}{2\pi}\Delta\gamma_L
\end{equation}is equal the change of the Zak (Berry) phase $\Delta\gamma_L = \gamma_L(\tau=T)-\gamma(\tau=0)$. Here, we introduced the current  $J(\tau)$ that drives the adiabatic change of the electric polarization. Note that the change of the polarization over one period $T$ is nothing other than the total charge transport averaged over space over one period $T$.\\

In a very similar way, we adapt these ideas to higher-order topological systems and HOSPTs. First, we introduce the particular choice of boundary conditions (Sec.~\ref{sec:boundaryconditions}) and the lattice Hamiltonian we work with (Sec.~\ref{sec:lattice_hamiltonian}). Then, similar to Resta, we introduce a family of Hamiltonians with flux $\Phi$ (Sec.~\ref{sec:flux_insertion}). If we set $\Phi=2\pi$ and use non-degenerate perturbation theory we find a relation between the higher-order Zak (Berry) phase $\gamma_i$ and the generator of the gauge transformation termed $\hat{X}_i$ (Sec.~\ref{sec:restas_construction}). In the same section we evaluate the time derivative of $\hat{X}_i$ (assuming an adiabatic change of parameters), which we relate to a current operator along the diagonals in Sec.~\ref{sec:currentOperator}. To connect the change of the higher-order Zak (Berry) phase over one period of time to a physical observable, we evaluate the adiabatic current, similar to Ref.~\cite{Thouless1983}, and find that the change of the higher-order Zak (Berry) phase is related to a charge transport per length (Sec.~\ref{sec:adiabatic_current}). In the last two sections we prove that the phase introduced in Sec.~\ref{sec:restas_construction} indeed corresponds to the higher-order Zak (Berry) phase (Sec.~\ref{sec:berryphase}) and that total charge transport per length is quantized in terms of the Chern number (Sec.~\ref{sec:chernnumber}).

\subsection{Corner Periodic Boundary Conditions}
\label{sec:boundaryconditions}
If we consider a square lattice called $\Lambda$ and choose the symmetry center as point of origin, then $\Lambda=\{(x,y):-D\leq x \leq D, -D\leq y \leq D\}$ with $D=(L-1)/2$ and CPBC are defined as follows (see also Fig.~\ref{sup_fig:flux}):
\begin{align}
    \label{sup_eq:CPBC}
    \hat{a}_{x,y+L}\equiv \hat{a}_{x,y}{\;}\text{if and only if}{\;}x=\pm \frac{(L-1)}{2}\nonumber\\
    \hat{a}_{x+L,y}\equiv \hat{a}_{x,y}{\;}\text{if and only if}{\;}y=\pm \frac{(L-1)}{2}.
\end{align}where $L$ is the number of lattice sites along the $x(y)$-direction. Unlike for systems with periodic boundary conditions, systems with CPBC have edges and only the corners are connected to each other. As we show below for the description of the 2D SL-BHM, CPBC are suitable for extending Resta's argument to higher-order topological states.
\subsection{Lattice Hamiltonian}
\label{sec:lattice_hamiltonian}
The lattice Hamiltonian we consider here is parametrized by a general parameter $\va\in\mathbb{R}^d$, where $d$ is the dimension of parameter space, and can be written as follows:
\begin{equation}
\label{sup_eq:hcpbc}
\h^{CPBC}_{\va} = \h^{OBC}_{\va_{OBC}} + \h^{C}_{\va_C}
\end{equation}where $\h^{OBC}_{\va_{OBC}}$ is the Hamiltonian with open boundaries and $\h^{C}_{\va_c}$ includes the links connecting the corners. 
Note that in general it's not necessary that both $\h^{OBC}_{\va_{OBC}}$ and $\h^{C}_{\va_C}$ depend on all parameters $\{\alpha_i\}$ with $\va=(\alpha_1,...,\alpha_d)$. To keep track of this, we introduced the parameters $\va_{OBC}$ and $\va_C$. For example, in the case of the model in the main text we have $\va=(t,U),{\;}\va_{OBC}=(t,U)$ and $\va_C=(t,0)$. Thus, $\h^{OBC}_{\va_{OBC}}$ depends on both parameters, while $\h^{C}_{\va_C}$ depends only on the hopping amplitude. Their definitions read:
\begin{align}
\label{sup_eq:SLBHMPBC}
    \hat{H}^{OBC}_{\va_{OBC}} &= -\Big[ \sum_{x=-D}^{D-1} \sum_{y=-D}^{D}(t(x)\, \hat{a}^{\dagger}_{x,y} \hat{a}_{x+1,y} + \mathrm{h.c.} ) + x \leftrightarrow y \Big]+ \frac{U}{2} \sum_{x,y=-D}^{D} \hat{n}_{x,y} (\hat{n}_{x,y}-1)\nonumber\\
    \h^{C}_{\va_{C}} & = -t\;( \hat{a}^{\dagger}_{c_{1}} \hat{a}_{c_{2}} +
    \hat{a}^{\dagger}_{c_{2}} \hat{a}_{c_{3}} 
    + \hat{a}^{\dagger}_{c_{3}} \hat{a}_{c_{4}} +
    \hat{a}^{\dagger}_{c_{4}} \hat{a}_{c_{1}}),
\end{align}where $D=(L-1)/2$ and $t(\zeta), \zeta \in \{x,y\}$ being
\begin{equation}
t(\zeta)  = \begin{cases}
1-t &\mathrm{for} \; \zeta \in  \{-D,-D+2,\dots,D-1\}\\
t & \mathrm{for}  \;\zeta \in  \{-D+1,-D+3,\dots,D-2\}
\end{cases}
\end{equation}and the corner coordinates are:
\begin{equation}
    c_1 = \left(-D,D\right),{\;}c_2 = \left(-D,-D\right),{\;}c_3 = \left(D,-D\right),{\;}c_4 = \left(D,D\right).
\end{equation} For what follows, we consider a general family of Hamiltonians $\h_{\va}^{CPBC} = \h^{OBC}_{\va_{OBC}} + \h^{C}_{\va_C}$ with $ \h^{OBC}_{\va_{OBC}}$ and $\h^{C}_{\va_C}$ defined in Eq.~\eqref{sup_eq:SLBHMPBC} where we allow for arbitrary on-site chemical potentials $\sim \sum_{x,y}\Delta_{x,y}\n_{x,y}$ in the definition of $\h^{OBC}_{\va_{OBC}}$, that might break global symmetries of $\h^{CPBC}_{\va}$. Clearly, such terms do not violate particle number conservation, which we assume throughout the following discussion. 
\subsection{Flux insertion}
\label{sec:flux_insertion}
In Fig.~\ref{fig:gauge_choice}(a) of the main text, the flux is inserted by twisting the hoppings connected to one of the four corners, i.e
\begin{equation}
    \label{sup_eq:hcpbc_theta}
    \h^{CPBC}_{\va, i} (\theta)=\h^{OBC}_{\va_{OBC}}+\hat{U}^{\dagger}_i(\theta)\h^{C}_{\va_C} \hat{U}_i(\theta)
\end{equation} for $i \in \{1, 2, 3, 4\}$. We emphasize that $\h^{OBC}_{\va_{OBC}}$ remains independent of $\theta$.

\begin{figure}[tbh!]
\includegraphics[width=0.4\textwidth]{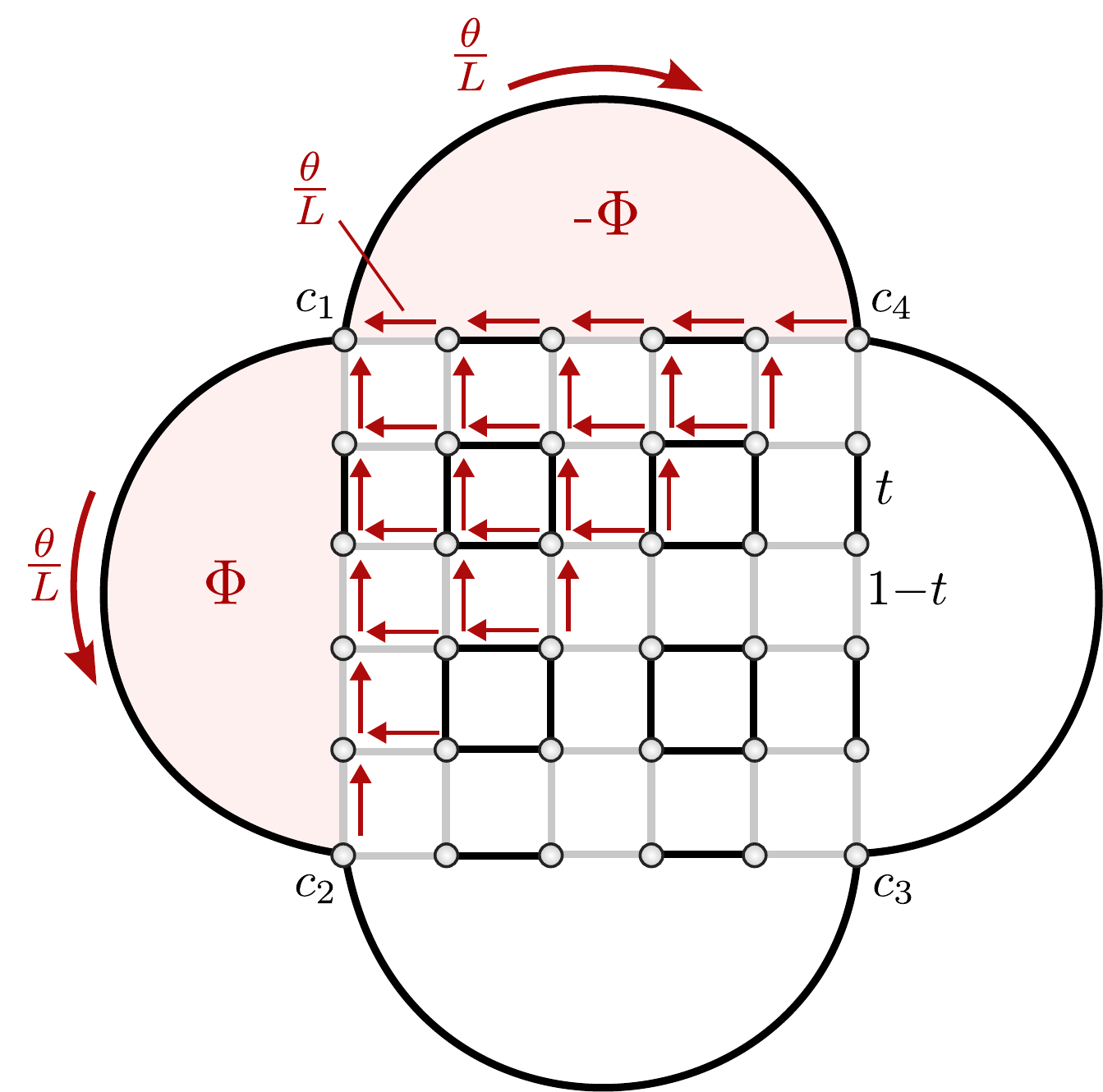}
\caption{In the new gauge the phase twist is distributed over all hoppings contained in upper triangle $T_1$. Each hopping is modulated by a phase factor with angle~$\theta/L$. The direction is such that a particle hopping from right (bottom) to left (top) acquires a phase factor $e^{i\theta/L}$. Inside the bulk there is no flux, but there is a flux $\Phi=\theta$ in the outer rings formed by CPBC. The gauge transformations for the other triangles can be obtained by applying $C_4$ symmetry on this configuration.}

\label{sup_fig:flux}
\end{figure}

However, similar to a one-dimensional system, the same flux $\Phi$ can result from different gauge choices. In one dimension, we either twist a single link or we distribute the twist over all links, i.e $t_{x,x+a}\to t_{x,x+a}e^{i\theta}$ or $t_{x,x+a}\to t_{x,x+a}e^{i\theta/L}$, where $t_{x,x+a}$ is the hopping amplitude between sites $(x,x+a)$ and $L$ is the total system size\footnote{This freedom to choose a gauge distinguishes the obtained Zak phase~\cite{Zak1989} from an ordinary Berry phase~\cite{Berry1984}.}. The second approach allows to apply perturbation theory in $\frac{\theta}{L}$ when $L \to \infty$, which is an essential part of Resta's argument~\cite{Resta1998}. Thus, we need to find another gauge transformation resulting in the same flux $\Phi$, that allows us to do perturbation theory in $1/L$ by distributing the twist of many links. A possible choice is shown in Fig.~\ref{sup_fig:flux}.

This new gauge choice can be obtained from the original one by applying a pure gauge transformation $\V_i(\theta)$ to the full Hamiltonian. This ensures that no flux is introduced in the bulk of the system. The gauge transformation here acts on $\h^{CPBC}_{\va}(\theta)$ given by  Eq.~\eqref{sup_eq:hcpbc_theta}:
\begin{equation}
    \label{sup_eq:hcpbc_theta2}
    \h^{CPBC}_{\va, i;\hat{V}}(\theta) = \V_i(\theta)\left(\h^{OBC}_{\va_{OBC}}+\hat{U}^{\dagger}_i(\theta)\h^{C}_{\va_C} \hat{U}_i(\theta)\right)\V_i^{\dagger}(\theta),
\end{equation}where we again defined four symmetry related gauge transformations $\V_i(\theta)$.
Note that the total flux $\Phi=\theta$ through the edge plaquettes is unchanged. 
Specifically, the gauge transformations $\V_i(\theta)$ distributing the twist over all links in the triangle $T_i$ are given by,
\begin{equation}
    \label{sup_eq:gaugetraf}
    \V_i({\theta}) = \exp\left(i\frac{\theta}{L}\hat{X}_i \right), {\;}\hat{X}_i = \sum_{x,y\in T_i}f_{x,y; i}\n_{x,y},
\end{equation} where $f_{x,y;i}$ is a function of lattice coordinates. The corresponding triangles, each spanned by three corners, are defined as
\begin{equation}
\label{sup_eq:triangles}
    T_1 = c_4\to c_1\to c_2,\; T_2 = c_1\to c_2\to c_3,\;T_3 = c_2\to c_3\to c_4,\; T_4 = c_3\to c_4\to c_1\;
\end{equation}where the labelling goes counterclockwise. For $i=1$ we obtain the model shown in Fig.~\ref{sup_fig:flux} and $f_{x,y;1}$ reads as follows:
\begin{equation}
    f_{x,y;1} =  (y-x),
\end{equation} Under $C_4$ symmetry the operators $\hat{X}_i$ and triangles $T_i$ are transformed into each other, $C_4:\hat{X}_i\to \hat{X}_{i+1}$ and $C_4:T_i\to \hat{T}_{i+1}$.

\subsection{Resta's construction}
\label{sec:restas_construction}
To follow the idea of Resta~\cite{Resta1998} we assume that the initial lattice Hamiltonian, i.e with no flux $\h^{CPBC}_{\va,i;V}(0)\equiv \h^{CPBC}_{\va} $, has a unique ground state $\h^{CPBC}_{\va}\ket{\Psi^0_{\va}}=E^{0}_{\va}\ket{\Psi^0_{\va}}$. Hence, if $\theta=2\pi$ (or equivalently $\Phi=2\pi$)\footnote{Note, compared to Ref.~\cite{Resta1998} we put the factor $1/L$ in the definition of the gauge transformation $\hat{V}_i(\theta)$.} then we find that
\begin{equation}
\label{sup_eq:hcpbc2pi}
  \h^{CPBC}_{\va,i;V}(2\pi)\,\V_i(2\pi)\,\ket{\Psi^0_{\va}}=E^{0}_{\va}\,\V_i(2\pi)\,\ket{\Psi^0_{\va}}.
\end{equation}In the following, we drop the subscript \glq$\hat{V}$\grq{} used in the definition of the Hamiltonian defined in Eq.~\eqref{sup_eq:hcpbc_theta2}. Whenever necessary, we explicitly state if we use Hamiltonian~\eqref{sup_eq:hcpbc_theta}. Using non-degenerate perturbation theory we find 
\begin{equation}
\label{sup_eq:pertT}
    \V_i(2\pi)\ket{\Psi^0_{\va}}=e^{i\tilde{\gamma}_{i}(\va)}\left(\ket{\Psi^0_{\va}}+\frac{2\pi}{L}\sum_{j>0}\ket{\Psi^{j}_{\va}}%
    \frac{\bra{\Psi^{j}_{\va}}\partial_{\tth} \h^{CPBC}_{\va,i}(\tth)|_{\tth=0}\ket{\Psi^{0}_{\va}}}{\left(E^0_{\va}-E^j_{\va}\right)}\right),
\end{equation}where the sum runs over all excited states. Moreover, we used that $\h^{CPBC}_{\va,i}(2\pi)=\h^{CPBC}_{\va,i}+\partial_{\tth}\h^{CPBC}_{\va,i}(\tth)|_{\tth=0}\frac{2\pi}{L}$, where we introduced the notation $\tth=\theta/L$\footnote{Since we put the factor $1/L$ in the defintion of the gauge transformation, taking the derivative w.r.t $\theta$ means we have to make use of the chain rule.}.  In Sec.~\ref{sec:berryphase} we show that the phase introduced in Eq.~\eqref{sup_eq:pertT} is related to the higher-order Zak (Berry) phase. More precisely, both phases coincide up to a shift which,in the presence of the $\mathbb{Z}_2$ symmetry introduced in Sec.~\ref{sec:quantization}, is constant. 
Due to CPBC, the expectation value of $\langle{\hat{X}_i\rangle}_{\va}=\braket{\Psi^0_{\va}|\hat{X}_i|\Psi^0_{\va}}$ is not well-defined. However, similar to the one dimensional case~\cite{Resta1998}, we can define the expectation value of a many-body operator as:
\begin{equation}
    \label{sup_eq:expQ}
    \langle{\hat{X}_i\rangle}_{\va}=\frac{L}{2\pi}\text{Im}\log\bra{\Psi^0_{\va}}e^{\frac{2\pi i}{L}\hat{X}_i}\ket{\Psi^0_{\va}}{\;}(\text{mod}{\;}L),
\end{equation}
which is well-defined modulo $L$. Using Eqs.~\eqref{sup_eq:pertT}~and~\eqref{sup_eq:expQ} we obtain
\begin{equation}
    \boxed{\langle{\hat{X_i}\rangle}_{\va}=\frac{L}{2\pi}\tilde{\gamma}_{i}(\va)}.
\label{sup_eq:polarization}
\end{equation}
Similarly to  Ref.~\cite{Resta1998}, we now want to relate Eq.~\eqref{sup_eq:polarization} to a physical observable. To do so, we assume an adiabatic parameter change $\va\to\va(\tau)$ where $\tau$ denotes time. Using the corresponding instantaneous eigenstates $\ket{\Psi^j_{\va(\tau)}}$ (short-hand we write $\ket{\Psi^j_{\va}}$) we can calculate the time derivative of Eq.~\eqref{sup_eq:expQ}
\begin{align}
    \label{sup_eq:dtQ1}
    \frac{\mathrm{d}}{\mathrm{d}\tau}\braket{\hat{X}_i}_{\va}=\frac{L}{2\pi}\text{Im}\left(\frac{\braket{\dot\Psi^0_{\va}|e^{\frac{2\pi i}{L}\hat{X}_i}|\Psi^0_{\va}}}{\braket{\Psi^0_{\va}|e^{\frac{2\pi i}{L}\hat{X}_i}|\Psi^0_{\va}}}+\frac{\braket{\Psi^0_{\va}|e^{\frac{2\pi i}{L}\hat{X}_i}|\dot\Psi^0_{\va}}}{\braket{\Psi^0_{\va}|e^{\frac{2\pi i}{L}\hat{X}_i}|\Psi^0_{\va}}}\right).
\end{align}Inserting the results of Eq.~\eqref{sup_eq:pertT} and keeping only first order terms we obtain
\begin{align}
\label{sup_eq:dtQ2}
    \frac{\mathrm{d}}{\mathrm{d}\tau}\braket{\hat{X}_i}_{\va} =& \sum_{j>0}\left[\braket{\dot\Psi^0_{\va}|\Psi^j_{\va}}\frac{\braket{\Psi^j_{\va}|(-i)\partial_{\tth}\h^{CPBC}_{\va,i}(\tth)|_{\tth=0}|\Psi^0_{\va}}}{E^0_{\va}-E^{j}_{\va}}+\braket{\Psi^j_{\va}|\dot\Psi^0_{\va}}\frac{\braket{\Psi^0_{\va}|i\partial_{\tth}\h^{CPBC}_{\va,i}(\tth)|_{\tth=0}|\Psi^j_{\va}}}{E^0_{\va}-E^{j}_{\va}}\right]\nonumber\\
    =& \sum_{j>0}\braket{\dot\Psi^0_{\va}|\Psi^j_{\va}}\frac{\braket{\Psi^j_{\va}|(-i)\partial_{\tth}\h^{CPBC}_{\va,i}(\tth)|_{\tth=0}|\Psi^0_{\va}}}{E^0_{\va}-E^{j}_{\va}}+\text{c.c.}
\end{align}where c.c. means complex conjugation. Here, similar to Resta~\cite{Resta1998}, we use that $\hat{H}_{\va}$ is time-reversal symmetric. As  we use spinless particles, this is just given by complex conjugation $\mathcal{T}=K$ with  $\mathcal{T}^{-1}\hat{a}_{x,y}\mathcal{T}=\hat{a}_{x,y}$, $\mathcal{T}^{-1}\hat{a}^\dagger_{x,y}\mathcal{T}=\hat{a}^\dagger_{x,y}$ and $\mathcal{T}^{-1} i \mathcal{T}=-i$. This symmetry guarantees that all instantaneous eigenstates can be chosen to be real and that the overlap $\braket{\dot\Psi^0_{\va}|\Psi^0_{\va}}=0$\footnote{The argument goes as follows: The ground state can always be chosen to be an eigenstate of $\mathcal{T}$ with $\mathcal{T}\ket{\Psi^0_{\va}}=\ket{\Psi^0_{\va}}$. From $\partial_\tau\braket{\Psi^0_{\va}|\Psi^0_{\va}}=0$ if follows that $\braket{\dot{\Psi}^0_{\va}|\Psi^0_{\va}}$ must be imaginary. However, if $\ket{\Psi^{0}_{\va}}$ is real, then also its time derivative is real. Thus, $\braket{\dot{\Psi}^0_{\va}|\Psi^0_{\va}}=0$. }. The imaginary unit $i$ appears because the derivative is of the form $\partial_{\tth}\h^{CPBC}_{\va,i}(\tth)|_{\tth=0}\sim i(\hat{A}^{\dagger}-\hat{A})$. Thus, taking its imaginary part corresponds to multiplying it with an additional factor of $(-i)$ or $i$ (for the complex conjugate part), respectively.\newline\\
\textcolor{gray}{To show this, let us define $a_{0j} = \braket{\dot\Psi^0_{\va}|\Psi^j_{\va}}$ and $b_{0j}=E^0_{\va}-E^{j}_{\va}$. Moreover we use that the derivative is of the form $\partial_{\tth}\h^{CPBC}_{\va,i}(\tth)|_{\tth=0}\sim i(\hat{A}^{\dagger}-\hat{A})$. Given this, we can write the imaginary part as follows:
\begin{align}
    &\text{Im}\left[\sum_{j>0}\frac{1}{b_{0j}}\left(a_{0j}\braket{\Psi^j_{\va}|\partial_{\tth}\h^{CPBC}_{\va,i}|_{\tth=0}|\Psi^0_{\va}}-\bar{a}_{0j}\braket{\Psi^0_{\va}|\partial_{\tth}\h^{CPBC}_{\va,i}|_{\tth=0}|\Psi^j_{\va}}\right)\right]\nonumber\\
    =&\text{Im}\left[\sum_{j>0}\frac{1}{b_{0j}}\left(ia_{0j}\left(\braket{\Psi^j_{\va}|A^{\dagger}|\Psi^0_{\va}}-\braket{\Psi^j_{\va}|A|\Psi^0_{\va}}\right)-i\bar{a}_{0j}\left(\braket{\Psi^0_{\va}|A^{\dagger}|\Psi^j_{\va}}-\braket{\Psi^0_{\va}|A|\Psi^j_{\va}}\right)\right)\right]\nonumber\\
    =&\text{Im}\left[\sum_{j>0}\frac{i}{b_{0j}}\left(a_{0j}\braket{\Psi^j_{\va}|A^{\dagger}|\Psi^0_{\va}}+\bar{a}_{0j}\braket{\Psi^0_{\va}|A|\Psi^j_{\va}}-a_{0j}\braket{\Psi^j_{\va}|A|\Psi^0_{\va}}-\bar{a}_{0j}\braket{\Psi^0_{\va}|A^{\dagger}|\Psi^j_{\va}}\right)\right]\nonumber\\
    =&\text{Im}\left[\sum_{j>0}\frac{i}{b_{0j}}\left(a_{0j}\braket{\Psi^j_{\va}|A^{\dagger}|\Psi^0_{\va}}+\bar{a}_{0j}\overline{\braket{\Psi^j_{\va}|A^{\dagger}|\Psi^0_{\va}}}-\left(a_{0j}\braket{\Psi^j_{\va}|A|\Psi^0_{\va}}+\bar{a}_{0j}\overline{\braket{\Psi^j_{\va}|A|\Psi^0_{\va}}}\right)\right)\right]\nonumber\\
    =&\hphantom{\text{Im}}\left[\sum_{j>0}\frac{1}{b_{0j}}\left(a_{0j}\braket{\Psi^j_{\va}|A^{\dagger}|\Psi^0_{\va}}+\bar{a}_{0j}\overline{\braket{\Psi^j_{\va}|A^{\dagger}|\Psi^0_{\va}}}-\left(a_{0j}\braket{\Psi^j_{\va}|A|\Psi^0_{\va}}+\bar{a}_{0j}\overline{\braket{\Psi^j_{\va}|A|\Psi^0_{\va}}}\right)\right)\right]
\end{align}Note, the relative minus sign in the first equation comes from the fact, that in the definition of the derivative in Eq.~\eqref{sup_eq:dtQ1} the second state is transformed by $\hat{V}^{\dagger}$ instead of $\hat{V}$, which is equivalent to an expansion around $\theta = -2\pi$.
Now we start with the other side:
\begin{align}
    &\left[\sum_{j>0}\frac{1}{b_{0j}}\left(a_{0j}\braket{\Psi^j_{\va}|(-i)\partial_{\tth}\h^{CPBC}_{\va,i}|_{\tth=0}|\Psi^0_{\va}}+\bar{a}_{0j}\braket{\Psi^0_{\va}|i\partial_{\tth}\h^{CPBC}_{\va,i}|_{\tth=0}|\Psi^j_{\va}}\right)\right]\nonumber\\
    =&\left[\sum_{j>0}\frac{1}{b_{0j}}\left(a_{0j}\left(\braket{\Psi^j_{\va}|A^{\dagger}|\Psi^0_{\va}}-\braket{\Psi^j_{\va}|A|\Psi^0_{\va}}\right)-\bar{a}_{0j}\left(\braket{\Psi^0_{\va}|A^{\dagger}|\Psi^j_{\va}}-\braket{\Psi^0_{\va}|A|\Psi^j_{\va}}\right)\right)\right]\nonumber\\
    =&\left[\sum_{j>0}\frac{1}{b_{0j}}\left(\left(a_{0j}\braket{\Psi^j_{\va}|A^{\dagger}|\Psi^0_{\va}}+\bar{a}_{0j}\braket{\Psi^0_{\va}|A|\Psi^j_{\va}}\right)-\left(a_{0j}\braket{\Psi^j_{\va}|A|\Psi^0_{\va}}+\bar{a}_{0j}\braket{\Psi^0_{\va}|A^{\dagger}|\Psi^j_{\va}}\right)\right)\right]\nonumber\\
    =&\left[\sum_{j>0}\frac{1}{b_{0j}}\left(a_{0j}\braket{\Psi^j_{\va}|A^{\dagger}|\Psi^0_{\va}}+\bar{a}_{0j}\overline{\braket{\Psi^j_{\va}|A^{\dagger}|\Psi^0_{\va}}}-\left(a_{0j}\braket{\Psi^j_{\va}|A|\Psi^0_{\va}}+\bar{a}_{0j}\overline{\braket{\Psi^j_{\va}|A|\Psi^0_{\va}}}\right)\right)\right]
\end{align}which is indeed the same as the above expression.
}
\subsection{Adiabatic current and total charge transport}
\label{sec:adiabatic_current}
Following Ref.~\cite{Niu1984}, we construct an expression for the adiabatic current and connect it to the time derivative in Eq.~\eqref{sup_eq:dtQ2}. Similar to our previous discussion, we use a short-hand notation, i.e. $\h^{CPBC}_{\va(\tau)}\equiv \h^{CPBC}_{\va}$. Note that after a period of $\tau=T$ the Hamiltonian returns to itself $\h^{CPBC}_{\va(\tau+T)}\equiv \h^{CPBC}_{\va(\tau)}.$ The density matrix associated with the adiabatic evolution is defined as
\begin{equation}
\label{sup_eq:densitym}
 \hat\rho_{\va}\equiv \hat\rho_{\va(\tau)}=\underbrace{\ket{\Psi^0_{\va(\tau)}}\bra{\Psi^0_{\va(\tau)}}}_{=\hat\rho^I_{\va(\tau)}}+\Delta\hat\rho(\tau)
\end{equation}where $\hat\rho^I_{\va(\tau)}\equiv \hat\rho^I_{\va}$ is the density matrix associated with instantaneous eigenstates. The time-evolution of the complete density matrix is governed by 
\begin{equation}
\label{sup_eq:timeevrho1}
 i\partial_\tau\hat\rho_{\va} = \left[\h^{CPBC}_{\va},\Delta\hat\rho(\tau)\right]. 
\end{equation}Dropping higher order terms such as $\partial_\tau\Delta\hat\rho(\tau)$~\cite{Kato1950adiabatic} we obtain for the instantaneous density matrix:
\begin{equation}
\label{sup_eq:timeevrho2}
    i\partial_\tau\hat\rho^I_{\va} \approx \left[\h^{CPBC}_{\va},\Delta\hat\rho(\tau)\right].
\end{equation}Further, we can write
\begin{align}
\label{sup_eq:0rhoij}
    \braket{\Psi^0_{\va}|i\left(\partial_\tau\hat\rho^I_{\va}\right)|\Psi^j_{\va}}=&i\underbrace{\partial_\tau \braket{\Psi^0_{\va}|\hat\rho^I_{\va}|\Psi^j_{\va}}}_{=0}-i\underbrace{\braket{\dot\Psi^0_{\va}|\hat\rho^I_{\va}|\Psi^j_{\va}}}_{=0}- i\braket{\Psi^0_{\va}|\hat\rho^I_{\va}|\dot\Psi^j_{\va}}\nonumber\\
    =&i\braket{\dot\Psi^0_{\va}|\Psi^j_{\va}}.
\end{align}Here, we used the following identities: First, the instantaneous eigenstates are orthogonal $\braket{\Psi^0_{\va}|\Psi^j_{\va}}=\delta_{0j}$. Second, the time derivative of $\partial_\tau \braket{\Psi^0_{\va}|\Psi^j_{\va}}=0$ vanishes and, thus, $\braket{\dot\Psi^0_{\va}|\Psi^j_{\va}}=-\braket{\Psi^0_{\va}|\dot\Psi^j_{\va}}$. Moreover, we find
\begin{equation}
\label{sup_eq:0hrhoj}
    \braket{\Psi^0_{\va}|\left[\h^{CPBC}_{\va},\Delta\hat\rho(\tau)\right]|\Psi^j_{\va}}=\left(E^0_{\va}-E^j_{\va}\right) \braket{\Psi^0_{\va}|\Delta\hat\rho(\tau)|\Psi^j_{\va}}.
\end{equation}Note, that the term proportional to $j=0$ vanishes. Inserting Eq.~\eqref{sup_eq:timeevrho2} into Eq.~\eqref{sup_eq:0hrhoj} and using the result of Eq.~\eqref{sup_eq:0rhoij} gives:
\begin{equation}
\label{sup_eq:0Deltarhoj}
    \braket{\Psi^0_{\va}|\Delta\hat\rho(\tau)|\Psi^j_{\va}}=i\frac{\braket{\dot\Psi^0_{\va}|\Psi^j_{\va}}}{E^0_{\va}-E^j_{\va}},{\;}j>0.
\end{equation}The total charge transport over one period $T$ averaged over space\footnote{The generator of the gauge transformation is linear in coordinates, which, as we shall see results in a charge transport along the diagonals, which thus gives rise to the factor $1/L.$} is defined as follows:
\begin{equation}
\label{sup_eq:totalchargecurrent}
    C = \frac{1}{L}\int^{T}_0\mathrm{d}\tau{\;}\text{tr}\left(\hat\rho_{\va}\hat J\right),
\end{equation}where $\hat J$ is the current operator, which, at this point is just a general operator and will be defined later. To evaluate the trace, we choose instantaneous eigenstates as our basis and replace $\hat\rho_{\va}$ with Eq.~\eqref{sup_eq:densitym}. Thus, the total charge transport is
\begin{align}
\label{sup_eq:chargecurrenteval}
     C =& \frac{1}{L}\int^{T}_0\mathrm{d}\tau{\;}\text{tr}\left(\hat{\rho}_{\va }\hat J\right)\nonumber\\
     =&\frac{1}{L}\int^{T}_0\mathrm{d}\tau\left[\left(1+\braket{\Delta\hat\rho}^{00}_{\va}\right)\braket{\hat J}^{00}_{\va}+\sum_{j>0}\braket{\Delta\hat\rho}^{0j}_{\va}\braket{\hat J}^{j0}_{\va}+\braket{\Delta\hat\rho}^{j0}_{\va}\braket{\hat J}^{0j}_{\va}\right].
\end{align}where $\braket{\cdot}^{jj'}_{\va}\equiv\braket{\Psi^j_{\va}|\cdot|\Psi^{j'}_{\va}}$. We used that there is no population of higher energy instantaneous eigenstates, i.e. $\braket{\Delta\hat\rho}_{jj'}=0$ for $j>0$ and $j'>0$ (adiabatic theorem~\cite{Kato1950adiabatic}). Since the current is odd under time-reversal symmetry, we have that $\braket{\hat J}_{00}=-\braket{\hat J}_{00}$. As the total charge current per length is real, the first term proportional to $\braket{\hat J}_{00}$ vanishes and we are left with (using Eq.~\eqref{sup_eq:0Deltarhoj})
\begin{align}
\label{sup_eq:chargecurrentprefinal}
C =& \frac{1}{L}\int^{T}_0\mathrm{d}\tau{\;}\text{tr}\left(\hat\rho_{\va}\hat J\right)\nonumber\\
=\frac{1}{L}&\sum_{j>0}\int^{T}_0\mathrm{d}\tau{\;}\left(\braket{\Delta\hat\rho}^{0j}_{\va}\braket{\hat J}^{j0}_{\va}+\braket{\Delta\hat\rho}^{j0}_{\va}\braket{\hat J}^{0j}_{\va}\right)\nonumber\\
\overset{\eqref{sup_eq:0Deltarhoj}}{=}&\frac{1}{L}\sum_{j>0}\int^{T}_0\mathrm{d}\tau{\;}\left(i\frac{\braket{\dot\Psi^0_{\va}|\Psi^j_{\va}}}{E^0_{\va}-E^j_{\va}}\braket{\hat J}^{j0}_{\va}+\text{c.c.}\right).
\end{align}The expression of the integral looks almost like the expression of Eq.~\eqref{sup_eq:dtQ2}. Thus, if we replace the current operator with $\hat J\to -\partial_{\tth}\h^{CPBC}_{\va,i}|_{\tth=0}$ and $C\to C_{c_i}$, we obtain
\begin{align}
\label{sup_eq:chargecurrentfinal}
    \boxed{C_{c_i}} =&\frac{1}{L} \sum_{j>0}\int^{T}_0\mathrm{d}\tau{\;}\left(\frac{\braket{\dot\Psi^0_{\va}|\Psi^j_{\va}}}{E^0_{\va}-E^j_{\va}}\braket{(-i)\partial_{\tth}\h^{CPBC}_{\va,i}|_{\tth=0}}^{j0}_{\va}+\text{c.c.}\right)\nonumber\\
    \overset{\eqref{sup_eq:dtQ2}}{=}&\frac{1}{L}\int^{T}_0\mathrm{d}\tau{\;}\frac{\mathrm{d}}{\mathrm{d}\tau}\braket{\hat{X}_i}_{\va} \nonumber\\
    \overset{\eqref{sup_eq:expQ}}{=}&\frac{\Delta\tilde{\gamma}_i}{2\pi}\nonumber\\
    {=}&\boxed{-\frac{\Delta \gamma_i}{2\pi}}
\end{align}
Given this replacement, we see that the total charge transport associated with the current defined by $\partial_{\tth}\h^{CPBC}_{\va,i}|_{\tth=0}$ as discussed in Sec.~\ref{sec:currentOperator}, (for hardcore bosons at half-filling) is proportional to the change of the higher-order Zak (Berry) phase $\Delta\gamma_i$ over one period (see Sec.~\ref{sec:berryphase}). In Sec.~\ref{sec:chernnumber} we prove that the change of the higher-order Zak (Berry) phase over one period $\Delta\gamma_i$ is quantized and equal to the Chern number. In the following section we show that this Chern number is proportional to the net charge $\Delta Q_{c_i}$ flowing through the corner $c_i$, thus the subscript on $C_{c_i}$.
\subsection{Current operator and corner charge}
\label{sec:currentOperator}
Now we need to justify that $\partial_{\tth}\h^{CPBC}_{\va,i}|_{\tth=0}$ indeed corresponds to a current. From the definition of $\h^{CPBC}_{\va,i}(\tth)$ (note that we defined $\tth=\theta/L$) in Eq.~\eqref{sup_eq:hcpbc_theta2} (see also Fig.~\ref{sup_fig:flux} for $i=1$) we derive\footnote{Note by applying $C_4$ symmetry we can generate all other derivatives $ \partial_{\tth}\h^{CPBC}_{\va,i}|_{\tth=0}$ and currents, respectively.} for $i=1$,
\begin{align}
\label{sup_eq:partialth}
   \partial_{\tth}\h^{CPBC}_{\va,1}|_{\tth=0}=&{i}\left[\sum_{y\in\Lambda}\sum_{x\in T^{'}_1}t(x)\left(\hat{a}^{\dagger}_{x+1,y}\hat{a}_{x,y}-\hat{a}^{\dagger}_{x,y}\hat{a}_{x+1,y}\right)-\sum_{x\in\Lambda}\sum_{y\in T^{'}_1}t(y)\left(\hat{a}^{\dagger}_{x,y+1}\hat{a}_{x,y}-\hat{a}^{\dagger}_{x,y}\hat{a}_{x,y+1}\right)\right]\nonumber\\
    \hphantom{=}& -it\left[\left(\hat{a}^{\dagger}_{c_4}\hat{a}_{c_1}-\hat{a}^{\dagger}_{c_1}\hat{a}_{c_4}\right)-\left(\hat{a}^{\dagger}_{c_1}\hat{a}_{c_2}-\hat{a}^{\dagger}_{c_2}\hat{a}_{c_1}\right) \right]\nonumber\\
    =&i\left(\Delta\hat{h}^{x,T^{'}_1}-\Delta\hat{h}^{y,T^{'}_1}\right)+i\left(-\Delta\hat{h}^{x,c_1}+\Delta\hat{h}^{y,c_1}\right)\nonumber\nonumber\\
    =&i\begin{pmatrix}\Delta\hat{h}^{x,T^{'}_1}\\ \Delta\hat{h}^{y,T^{'}_1}\end{pmatrix}\cdot 
    \begin{pmatrix}1\\-1\end{pmatrix} +i\begin{pmatrix}\Delta\hat{h}^{x,c_1}\\ \Delta\hat{h}^{y,c_1}\end{pmatrix}\cdot 
    \begin{pmatrix}-1\\1\end{pmatrix}\nonumber\\
    =& \hat{J}^{T^{'}_1}(\searrow)+\hat{J}^{c_1}(\nwarrow)
\end{align}
where $T^{'}_1$ is defined such that the sums in the above equation do not contain the corner connecting hopping terms. The operator $\Delta \hat{h}^{x(y), T^{'}_1}$ defines the net particles hopping along the $x(y)$-direction in $T^{'}_1$ and the operator $\Delta\hat{h}^{x(y), c_1}$ defines a similar expression for the corner $c_1$ only. Multiplying those terms with the imaginary unit gives indeed a current. Note that the sign of each term has been chosen such that a positive expectation value of $i\Delta h^{\nu,T^{'}_1(c_1)}$ with $\nu\in\{x,y\}$ means a current along $x(y)$. The final result shows that a non-zero expectation value of the derivative $ \partial_{\tth}\h^{CPBC}_{\va,1}|_{\tth=0}$ is equal to the projection of the current along the diagonal in the triangle $T_1$. Thus, this expectation value measures the net charge $\Delta Q_{c_1}$ that passes the corner over one period of time $T$.

Thus, from Eq.~\eqref{sup_eq:chargecurrentfinal} we conclude that the net charge passing the corner during one period of time is equal to the change of the higher-order Zak (Berry) phase (for hardcore bosons at half-filling) obtained from gauging the corner $c_i$ over one period of time $T$ 
\begin{equation}
\label{sup_eq:ChargeBerry}
    \boxed{-\frac{\Delta\gamma_i}{2\pi} = \Delta Q_{c_i}.}
\end{equation}\newline\\
\textcolor{gray}{
To make this more clear, let us consider the one-dimensional analogue. The hopping part of such a Hamiltonian including the twists is given by (note that $\tth=\theta/L)$
\begin{equation}
    \h^{hop}_{\va}(\tth) = -\left[\sum^{L}_{x=1}t(x)e^{i\tth}\hat{a}^{\dagger}_{x+1}\hat{a}_x+\text{h.c.}\right]
\end{equation}and the corresponding derivative reads:
\begin{align}
   \partial_{\tth}\h^{hop}_{\va}|_{\tth=0}=&-{i}\left[\sum^{L-1}_{x=1}t(x)\left(\hat{a}^{\dagger}_{x+1}\hat{a}_{x}-\hat{a}^{\dagger}_{x}\hat{a}_{x+1}\right)\right]- it\left(\hat{a}^{\dagger}_{1}\hat{a}_{L}-\hat{a}^{\dagger}_{L}\hat{a}_{1}\right)\nonumber\\
    =&-i\Delta\hat{h}^{x,bulk}+i\Delta\hat{h}^{x,edge}\nonumber\\
    =&i\begin{pmatrix}\Delta\hat{h}^{x,bulk}\\0\end{pmatrix}\cdot 
    \begin{pmatrix}-1\\0\end{pmatrix} +i\begin{pmatrix}\Delta\hat{h}^{x,edge}\\ 0\end{pmatrix}\cdot 
    \begin{pmatrix}1\\0\end{pmatrix}\nonumber\\
    =& \hat{J}^{bulk}(\leftarrow)+\hat{J}^{edge}(\rightarrow).
\end{align} If a particle hops from site $L\to 1$, it hops along the negative x-direction which introduces the additional minus sign. The differences are again defined such that a positive expectation value means a current along the positive $x$-axis. From this, we see that the Chern number measures the net charge flowing through the right/left edge. }

\subsection{The relation to the higher-order Zak (Berry) phase}
\label{sec:berryphase}
This section is dedicated to showing that the phase introduced in Eq.~\eqref{sup_eq:pertT} corresponds to the higher-order Zak (Berry) phase defined in Sec. \ref{sec:quantization}.
To this end, we first evaluate the ground states of $\h^{CPBC}_{\va,i;V}(\theta)$ using non-degenerate perturbation theory up to first order, similar to Eq.~\eqref{sup_eq:pertT},
\begin{equation}
\label{sup_eq:expansionofstates}
    \ket{\Psi^0_{\va,i;V}(m\Delta\theta)} = e^{i\phi_m}\left(\ket{\Psi^0_{\alpha}}+\frac{m\Delta\theta}{L}\sum_{j>0}\ket{\Psi^{j}_{\va}}%
    \frac{\bra{\Psi^{j}_{\va}}\partial_{\tth} \h^{CPBC}_{\va,i;V}(\tth)|_{\tth=0}\ket{\Psi^{0}_{\va}}}{\left(E^0_{\va}-E^j_{\va}\right)}\right),\quad m\Delta\theta=m\frac{2\pi}{n}
\end{equation}
where $m\leq n$ and $\tth = \theta/L$. As a next step, we evaluate the overlap of two states differing by $\Delta\theta$
\begin{align}
\label{sup_eq:overlap}
    \braket{\Psi^0_{\va,i;V}(m\Delta\theta)|\Psi^0_{\va,i;V}((m+1)\Delta\theta)}=e^{i(\phi_{m+1}-\phi_m)}\left(1+\frac{m(m+1)\Delta\theta^2}{L^2}\sum_{j}\frac{\left|\bra{\Psi^{j}_{\va}}\partial_{\tth} \h^{CPBC}_{\va,i;V}(\tth)|_{\tth=0}\ket{\Psi^{0}_{\va}}\right|^2}{\left(E^0_{\va}-E^j_{\va}\right)^2}\right),
\end{align}
where the term inside the brackets is real. If we went to higher-order perturbation theory, then we would also get imaginary terms from evaluating overlaps. However, the first imaginary term is proportional to $L^{-3}$, which results from the cross term of first and second order perturbation theory.
\\

\color{gray}
{Indeed, time-reversal symmetry enforces the expectation value of $\bra{\Psi^{j}_{\va}}\partial_{\tth} \h^{CPBC}_{\va,i;V}(\tth)|_{\tth=0}\ket{\Psi^{l}_{\va}}$ to be purely imaginary. For a time-reversal symmetric system, i.e., invariance under complex conjugation $\mathcal{T}=K$ we can choose real eigenstates such that the following holds:
\begin{equation}
    \bra{\Psi^{j}_{\va}}\partial_{\tth} \h^{CPBC}_{\va,i;V}(\tth)|_{\tth=0}\ket{\Psi^{l}_{\va}}=\bra{K\Psi^{j}_{\va}}\partial_{\tth} \h^{CPBC}_{\va,i;V}(\tth)|_{\tth=0}\ket{K\Psi^{l}_{\va}}=-\bra{\Psi^{l}_{\va}}\partial_{\tth} \h^{CPBC}_{\va,i;V}(\tth)|_{\tth=0}\ket{\Psi^{j}_{\va}}.
\end{equation}Here we used that $\partial_{\tth} \h^{CPBC}_{\va,i;V}(\tth)|_{\tth=0}\sim i (A^{\dagger}+A)$ with $K A K=A$. Second, we use that $\partial_{\tth} \h^{CPBC}_{\va,i;V}(\tth)|_{\tth=0}$ is a hermitian operator, 
\begin{equation}
    \overline{\bra{\Psi^{l}_{\va}}\partial_{\tth} \h^{CPBC}_{\va,i;V}(\tth)|_{\tth=0}\ket{\Psi^{j}_{\va}}}=\bra{\Psi^{j}_{\va}}\partial_{\tth} \h^{CPBC}_{\va,i;V}(\tth)|_{\tth=0}\ket{\Psi^{l}_{\va}}.
\end{equation}Thus, equating the above equations gives
\begin{equation}
    \overline{\bra{\Psi^{l}_{\va}}\partial_{\tth} \h^{CPBC}_{\va,i;V}(\tth)|_{\tth=0}\ket{\Psi^{j}_{\va}}} = -\bra{\Psi^{l}_{\va}}\partial_{\tth} \h^{CPBC}_{\va,i;V}(\tth)|_{\tth=0}\ket{\Psi^{j}_{\va}}
\end{equation} and hence $\bra{\Psi^{l}_{\va}}\partial_{\tth} \h^{CPBC}_{\va,i;V}(\tth)|_{\tth=0}\ket{\Psi^{j}_{\va}}=i\left|\bra{\Psi^{l}_{\va}}\partial_{\tth} \h^{CPBC}_{\va,i;V}(\tth)|_{\tth=0}\ket{\Psi^{j}_{\va}}\right|$. This implies that a product of three such terms is purely imaginary. The first time such term appears is if we have a cross product of a first order and second order term, which would be of third order in the coupling strength $\Rightarrow$ $L^{-3}$.}
\\
\color{black}

As a third step, we use that in the limit $n\to\infty$ the higher-order Zak (Berry) phase can be written in a discretized form~\cite{resta2007theory} given by\footnote{Note that $\ket{\Psi_{\va,i;V}(0)}=\ket{\Psi_{\va,i}(0)}$.}
\begin{equation}
\label{sup_eq:Berrydiscrete}
    \gamma_{i;V}(\va) =-\lim_{n\to\infty}\text{Im}\log \braket{\Psi_{\va,i}(0)|\Psi_{\va,i;V}(\Delta\theta)}\braket{\Psi_{\va,i;V}(\Delta\theta)|\Psi_{\va,i;V}(2\Delta\theta)}\cdots\braket{\Psi_{\va,i;V}(n\Delta\theta)|\hat{V}_i(2\pi)|\Psi_{\va,i}(0)}.
\end{equation}
To distinguish the higher-order Zak (Berry) phases obtained from different Hamiltonians we added a subscript here.
We also have to include the gauge transformation in the last step since after a $2\pi $ flux insertion the Hamiltonian $\h^{CPBC}_{\va,i;V}(\theta)$ does not return to itself; instead $\h^{CPBC}_{\va,i;V}(\theta+2\pi)$=$\hat{V}_i(2\pi)\h^{CPBC}_{\va,i;V}(\theta)\hat{V}_i^{\dagger}(2\pi)$.\\
\color{gray}
The discretized formula, Eq.~\eqref{sup_eq:Berrydiscrete} (assuming a $2\pi$ periodic Hamiltonian), can be obtained as follows: First, we rewrite the overlap between two states differing by an angle $\Delta\theta$ as
\begin{equation}
    \braket{\Psi_{\va,i}(m\Delta\theta)|\Psi_{\va,i}(m+1)\Delta\theta)}=1-i\left(i\braket{{\Psi_{\va,i}(m\Delta\theta)}|\partial_{\theta}|\Psi_{\va,i}(m\Delta\theta)}\Delta\theta\right),
\end{equation}which in the limit of $m\to\infty$ becomes
\begin{equation}
    \braket{\Psi_{\va,i}(m\Delta\theta)|\Psi_{\va,i}(m+1)\Delta\theta)} \overset{m\to\infty}{=}\exp\left[-i\left(i\braket{{\Psi_{\va,i}(m\Delta\theta)}|\partial_{\theta}|\Psi_{\va,i}(m\Delta\theta)}\Delta\theta\right)\right].
\end{equation}Second, taking the infinite product of such an expression is equivalent to the line integral of the exponent
\begin{equation}
-\text{Im}\log\prod^{\infty}_{m=1} \exp\left[-i\left(i\braket{{\Psi_{\va,i}(m\Delta\theta)}|\partial_{\theta}|\Psi_{\va,i}(m\Delta\theta)}\Delta\theta\right)\right]=\int^{2\pi}_0\mathrm{d}\theta{\;}i\braket{{\Psi_{\va,i}(\theta)}|\partial_{\theta}|\Psi_{\va,i}(\theta)}
\end{equation}which is nothing other than the definition of the higher-order Zak (Berry) phase $\gamma_i(\va)$.\\
\color{black}
If we insert the result of Eq.~\eqref{sup_eq:overlap} into the discretized version of the higher-order Zak (Berry) phase, Eq.~\eqref{sup_eq:Berrydiscrete}, we obtain
\begin{align}
\label{sup_eq:gVtg}
    \gamma_{i;V}(\va) \overset{L\to\infty}{=} - \mathrm{Im}\log\braket{\Psi^0_{\va,i}|\hat{V}_{i}(2\pi)|\Psi^{0}_{\va,i}}=-\tilde{\gamma}_{i}(\va).
\end{align}
Before we proceed let us briefly comment on the minus sign. For the higher-order Zak (Berry) phase we evaluated the phase being tuned from a state at $2\pi$ to a state at $0$ flux, while for $\tilde{\gamma}_i(\va)$ it is the opposite.

Now we want to relate the higher-order Zak (Berry) phases obtained from different Hamiltonians defined in Eq.~\eqref{sup_eq:hcpbc_theta}. By construction, different ground states are related as follows: $\ket{\Psi_{\va,i;V}(\theta)}=\hat{V}_i(\theta)\ket{\Psi_{\va,i}(\theta)}$. Thus, we get:
\begin{align}
\label{sup_eq:gVg}
     \gamma_{i;V}(\va) =& i \int^{2\pi}_0\mathrm{d}\theta\;\braket{\Psi_{\va,i;V}(\theta)|\partial_\theta|\Psi_{\va,i;V}(\theta)}\nonumber\\
     =&\gamma_{i}(\va) - \int^{2\pi}_0\mathrm{d}\theta\;\braket{\Psi_{\va,i}(\theta)|\frac{\hat{X}_i}{L}|\Psi_{\va,i}(\theta)}\nonumber\\
     =&\gamma_i(\va) -\bar{X}_i(\va).
\end{align}Hence, the higher-order Zak (Berry) phases obtained from different Hamiltonians differ by the averaged expectation value of the generator of the gauge transformation $V_{i}(\theta)$. Similar to the quantization of the higher-order Zak (Berry) phase (Sec.~\ref{sec:quantization}), we obtain (assuming $\mathbb{Z}_2$ symmetry):
\begin{align}
\label{sup_eq:Xnicely}
    \int^{2\pi}_0\mathrm{d}\theta\;\braket{\Psi_{\va,i}(\theta)|\frac{\hat{X}_i}{L}|\Psi_{\va,i}(\theta)} =& -\int^{2\pi}_0\mathrm{d}\theta\;\braket{\Psi_{\va,i}(-\theta)|\frac{\hat{X}_i}{L}|\Psi_{\va,i}(-\theta)}+\frac{2\pi}{L}\sum_{x,y\in T_i}f_{x,y;i}\nonumber\\
    =&-\int^{2\pi}_0\mathrm{d}\theta\;\braket{\Psi_{\va,i}(\theta)|\frac{\hat{X}_i}{L}|\Psi_{\va,i}(\theta)}+\frac{2\pi}{L}\sum_{x,y\in T_i}f_{x,y;i}\nonumber\\
    \Rightarrow\hphantom{=}&  \bar{X}_i(\va) = \frac{\pi}{L}\sum_{x,y\in T_i}f_{x,y;i}.
\end{align}Consequently, the shift between the higher-order Zak (Berry) phases is given by the average filling multiplied with $f_{x,y;i}$ in this region (given that $\mathbb{Z}_2$ symmetry is preserved), which is independent of the parameters in the Hamiltonian and,thus, constant. Otherwise, there is no reason for this shift to be quantized, however, as we shall see for evaluating the Chern number it is irrelevant.
From this, using Eq.~\eqref{sup_eq:gVtg}, we find that:
\begin{equation}
\label{sup_eq:gtg}
    \tilde{\gamma}_i(\va) = \bar{X}_i(\va)-\gamma_{i}(\va).
\end{equation}
If we, in the symmetric case, subtract the average charge in the definition of $\hat{X}_i$, then both phases coincide up to a minus sign
\begin{equation}
    \gamma_{i}(\va) =  -\mathrm{Im}\log\braket{\Psi^0_{\va,i}|\hat{V}'_{i}(2\pi)|\Psi^{0}_{\va,i}} = -\tilde{\gamma}'_{i}(\va)
\end{equation}where the superscript signals that we subtracted the average filling\footnote{This is similar to the one-dimensional case, where the Zak phase and the polarization defined in Ref.~\cite{Resta1998} differ by a constant shift if one does not subtract the average charge.}.
However, note that this is actually not necessary because at the end of the day only differences in the higher-order Zak (Berry) phases are accessible \cite{Atala2013}, which are unchanged under constant shifts. Moreover, at the symmetric points both phases $\gamma_{i}(\va)$ and $\tilde{\gamma}'_{i}(\va)$ are quantized to be $0,\pi\;\mathrm{mod}\;2\pi$ such that the global minus sign is irrelevant\footnote{Note, if we define $\tilde{\gamma}_{i}(\va)$ with respect to $\hat{V}'_{i}(2\pi)$ then it is easily shown --- without a relation to $\gamma_i(\va)$ --- that it is quantized to values $0,\pi$ at $\mathbb{Z}_2$ symmetric points.} for characterizing individual phases. For the Thouless pumps, however, this sign is important.
\textit{Thouless pump}. For the pumping procedure we explicitly break all symmetries that quantizes the higher-order Zak (Berry) phases. This naturally implies that the shift along such paths is no longer constant. However, for the loop integral over one period only the multi-valued part of $\gamma_{i;V}(\va)$ contributes non-trivially, which in Eq.~\eqref{sup_eq:gVg} is given by $\gamma_{i}(\va)$. The shift $\bar{X}_i(\va)$ is gauge-invariant and periodic in $\tau\to \tau+T$. Thus, the integral of $\partial_\tau \bar{X}_i(\va)$ over one-period vanishes. Consequently, we find that the average charge transport over one period defined in Eq.~\eqref{sup_eq:chargecurrentfinal} indeed corresponds to the change of the higher-order Zak (Berry) phase over one period. 

\subsection{Chern number}
\label{sec:chernnumber}

In this section we show that the total charge transport through the corner is related to the Chern number. Therefore, we go back to the second line of Eq.~\eqref{sup_eq:chargecurrentfinal} and substitute the expectation value $\braket{\hat{X}_i}_{\va}$ with Eq.~\eqref{sup_eq:polarization}. We obtain:
\begin{align}
     \frac{1}{L}\int^{T}_0\mathrm{d}\tau{\;}\partial_\tau\braket{\hat{X}_i}_{\va}=&\frac{1}{2\pi} \int^{T}_0\mathrm{d}\tau{\;}\partial_\tau\tilde{\gamma}_{i}(\va)\nonumber\\
    =-&\frac{1}{2\pi} \int^{T}_0\mathrm{d}\tau{\;}\partial_\tau{\gamma}_{i}(\va)\nonumber\\
    =-& \frac{i}{2\pi} \int^{T}_0\mathrm{d}\tau{\;}\partial_\tau \int^{2\pi}_0\mathrm{d}\theta{\;}\braket{\Psi_{\va,i}(\theta)|\partial_\theta|\Psi_{\va,i}(\theta)}\nonumber\\
    =&-\frac{i}{2\pi}\int^{T}_0\mathrm{d}\tau{\;}\int^{2\pi}_0\mathrm{d}\theta{\;}\left(\braket{\partial_\tau\Psi_{\va,i}(\theta)|\partial_\theta\Psi_{\va,i}(\theta)}-\braket{\partial_\theta\Psi_{\va,i}(\theta)|\partial_\tau\Psi_{\va,i}(\theta)}\right)\nonumber\\
    =&-\mathcal{C}_{i},
\end{align}where we used that $\braket{\partial_\theta\Psi_{\va,i}(\theta)|\Psi_{\va,i}(\theta)}=-\braket{\Psi_{\va,i}(\theta)|\partial_\theta\Psi_{\va,i}(\theta)}$ and Eq.~\eqref{sup_eq:gtg} (recall that the loop integral over the shift vanishes). The last identity is equal to the Chern number, a result that was obtaind by Thouless~et.~al in Ref.~\cite{Thouless1982Chern}. Thus, we find that the change of the higher-order Zak (Berry) phase over one period $T$ is quantized:
\begin{equation}
\label{sup_eq:Chernberry}
    \boxed{\mathcal{C}_i=\frac{\Delta\gamma_i}{2\pi}}.
\end{equation}which using Eq.~\eqref{sup_eq:ChargeBerry} gives us the quantization of the corner charge:
\begin{equation}
\label{sup_eq:Cherncornercharge}
    \boxed{\mathcal{C}_i=-\Delta Q_{c_i}}.
\end{equation}
\section{Finite Gap}
In this section we review why the bulk energy gap does not close when computing the Zak (Berry) phases.
Before we present numerical calculations we briefly recap a perturbation theoretic argument why the flux insertion, as it is introduced in the current work, cannot close the bulk gap in the thermodynamic limit. 
In section \ref{sec:flux_insertion} we showed that phase twist $e^{i\theta}$ along the corner connecting links can be distributed to parts of the bulk such that each hopping carries a complex phase of max. $\theta/L$. At the same time, the total flux inside each plaquette remains zero, such that there is zero flux in the bulk. Since the Hamiltonian is modified locally only in powers of $1/L$, we can apply perturbation theory (assuming that the initial system has a finite bulk gap, perturbation theory is applied to each energy level separately). Hence, the initial bulk gap is modified by the terms that are proportional to powers of $1/L$ multiplied by finite corrections if there are no degeneracies in the low-lying energy levels. For larger system sizes $L$ the perturbations become weaker and thus the changes of the bulk according to the phase twist decrease, which implies that the bulk gap cannot close. 

~
\begin{figure}[tbh!]
\includegraphics[width=\textwidth]{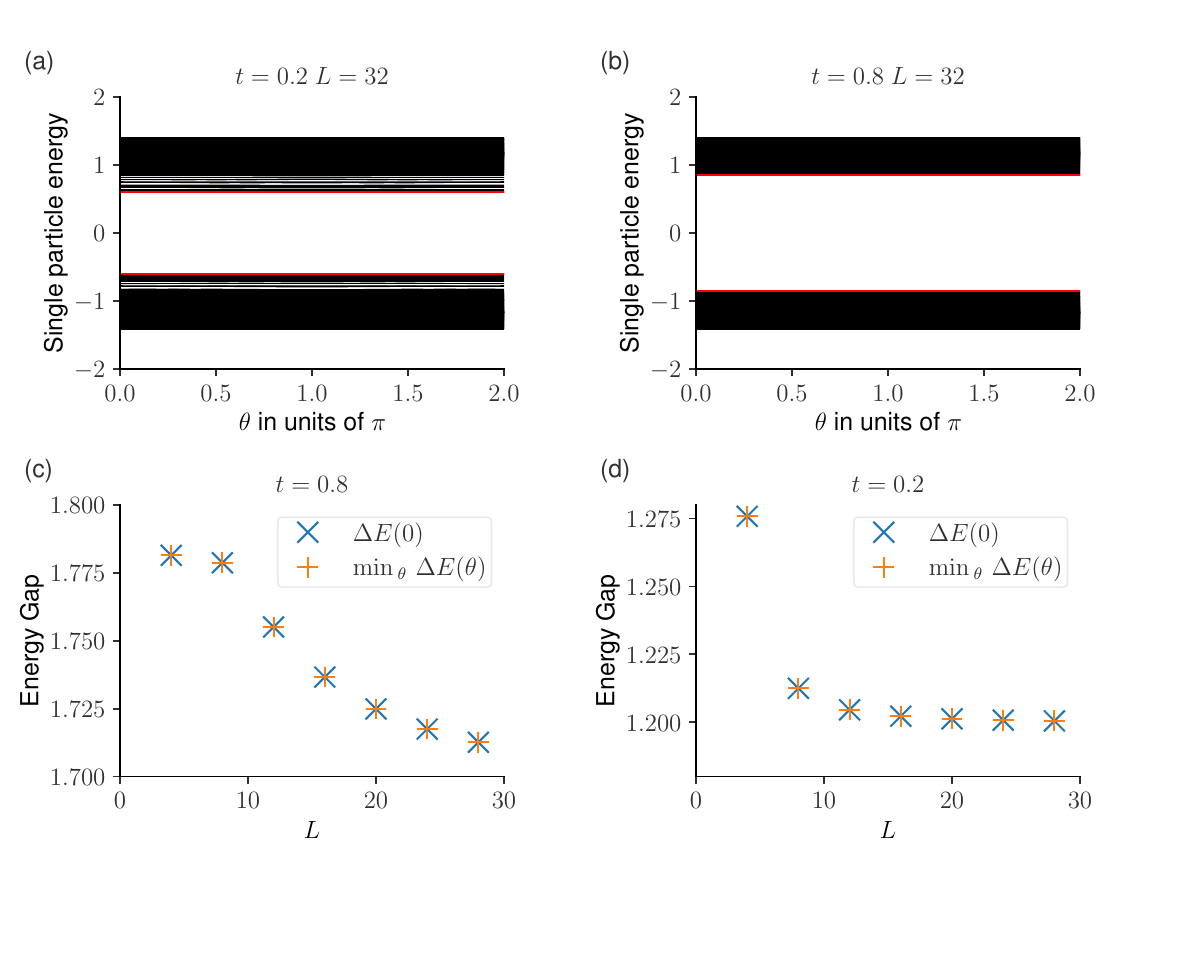}
    
     \caption{\textbf{(a)-(b) BBH Energy spectrum over $\theta$ for $L=32$. }Shown is the full energy spectrum for the BBH model with CPBC over $\theta$, that introduces flux in the super-cells, in the trivial $t=0.2$ (a) and non-trivial $t=0.8$ (b) case. The $N$-th and $(N+1)$-th state, that determine the gap between the ground and first excited state are marked in red. Here, $N=L\times L/2$ corresponds to the number of particles at half-filling. The introduced flux has no visible influence on the spectrum. \textbf{(c)-(d) Energy gap over length.} Shown are the bulk energy gaps $\Delta E = E_e-E_g$ between the ground and first excited state at different lengths in the trivial (c) and non-trivial (d) case. Marked in blue is the energy gap without flux insertion. Orange plusses mark the numerically minimal gap that was found when varying $\theta$. For all computed lengths the gaps have approximately the same value. }
     \label{fig:energylength}
 \end{figure}

For numerical evidence, we compute the influence of the flux insertion on the energy of the ground and first excited state of the Benalcazar-Bernevig-Hughes (BBH)  model \cite{Benalcazar2017}. The BBH model in two dimensions is the free fermion analogue of the SL-BH model, which  reduces to the BBH model for $U \to \infty$ and $t \in \{0, 1\}$ \cite{Bibo2020}. At half-filling, i.e.\ number of particles $N=L\times L/2$  it is known that the BBH model preserves the energy gap ( when $t\neq 0.5$). Hence, we only have to show that the insertion of the flux does not close this gap. We use corner periodic boundary conditions and insert the flux in the superplaquettes as described in the main text of our manuscript.
As shown earlier, there are four possible gauge transformations that can introduce the flux, but as they are connected by $C_4$ symmetry here we only consider the first one.

~

Figure~\ref{fig:energylength} (a-b) shows the energy spectrum of this model for a system of length $L=32$ for different values of $\theta$ for the two cases of $t=0.2$ (~\ref{fig:energylength}(a)) and $t=0.8$ (~\ref{fig:energylength}(b)). The $N$-th and $(N+1)$-th state, that determine the gap between the ground and first excited state are marked in red. The plot exemplifies that the influence of $\theta$ on the overall energy spectrum is negligible. To show that this is the case also for different lengths we plot the energy gap without flux $\Delta E(0)$ and the minimal energy gap $\min_{} {}_\theta \Delta E(\theta)$ over different system lengths. This verifies that the influence of $\theta$ on the energy gap is vanishing for all considered system sizes.

\putbib[references_supplement]
\end{bibunit}


\begin{thebibliography}{54}
\bibitem [{sig()}]{signnote}%
  \BibitemOpen
  \href@noop {} {}\bibinfo {note} {Note that in the numerical calculations we
  defined the Berry Phase positively, such that the minus sign
  cancels.}\BibitemShut {Stop}%
\end{thebibliography}
\end{document}